\documentclass[aps,preprint,superscriptaddress,showpacs,floatfix]{revtex4}
\usepackage{graphicx}
\usepackage{amsmath}
\usepackage{wrapfig}

\begin{document}

\title{Dengue epidemics and human mobility}
\author{D.H. Barmak, C.O. Dorso, M. Otero, H.G. Solari}
\affiliation{Departamento de F\'{\i}sica, Facultad de Ciencias Exactas y Naturales,
Universidad de Buenos Aires, Pabell\'on $I$, Ciudad Universitaria, Nu\~{n}%
ez, $1428$,\\
Buenos Aires, Argentina.}
\date{\today}
\begin{abstract}
In this work we explore the effects of human mobility on the dispersion of a vector borne disease. We combine an already presented stochastic model for dengue with a simple representation 
of the daily motion of humans on a schematic city of 20x20 blocks with 100 inhabitants in each block. The pattern of motion of the individuals is described in terms of complex networks 
in which links connect different blocks and the link length distribution is in accordance with recent findings on human mobility. It is shown that human mobility can turn out to be the 
main driving force of the disease dispersal.
\end{abstract}

\maketitle

\section{Introduction}

Dengue fever is a vector borne disease produced by a \emph{flavivirus} of
the family \emph{flaviviridae} \cite{gubl98}. The main vectors of dengue are 
\emph{Aedes aegypti} and \emph{Aedes albopictus}.

Dengue fever epidemics have been known for more then two centuries as benign
fever epidemics appearing at intervals of one to several decades.
Recently the number of countries affected by this epidemic
has increased and severe forms have become more frequent.

The research aimed at producing dengue models for public policy use began
with Newton and Reiter \cite{newt92} who introduced a minimal model for
dengue in the form of a set of Ordinary Differential Equations (ODE) for the
human population disaggregated in Susceptible, Exposed, Infected and
Recovered compartments. The mosquitoes population was not modeled in this
early work. A different starting point was taken by Focks et al. \cite%
{fock93a,fock93b} that began by describing mosquitoes populations in a
computer framework named Dynamic Table Model where later the human
population (as well as the disease) was introduced \cite{fock95}.
 
Newton and Reiter's model (\textbf{NR}) favors economy of resources and
mathematical accessibility, in contrast, Fock's model emphasize realism,
these models represent in Dengue two contrasting compromises in the standard
trade off in modeling. A third starting point has been recently added.
Otero and Solari (\textbf{OS}) developed a dengue model \cite%
{oter10} which includes the evolution of the mosquitoes population \cite%
{oter06, oter08} and is spatially explicit. This last model is somewhat in
between Fock's and NR as it is formulated as a state-dependent Poisson model
with exponentially distributed times.

ODE models have received
most of the attention. Some of the works explore: variability of vector
population \cite{este98}, human population \cite{este99}, the effects of
hypothetical vertical transmission of Dengue in vectors \cite{este00},
seasonality \cite{bart02}, age structure \cite{pong03} as well as incomplete
gamma distributions for the incubation and infectious times \cite{chow07}.
Comparison with real epidemics has shown that there is a need to consider
the spatial heterogeneity as well \cite{favi05}.

In a previous work \cite{oter11} we have developed a
dengue model which includes the evolution of the mosquitoes population and 
is spatially explicit. In that work the spatial spread of the infection was
driven by the flights of the mosquitoes that gave rise to a diffusion
process. In it we analyzed the evolution of dengue infection in a
city of 20X20 blocks with 100 individuals in each one, this population
was fixed throughout the calculation and no mobility of the
individuals was allowed.

As such, in that model, the spatial evolution of the dengue infection was
only driven by the flight of mosquitoes as the mobility of humans was not
included. It is usually recognized that human mobility is not only necessary
to be included in human infection spread models, but that it might be the
main source of the dynamics behind spatiotemporal phenomena on geographic
scales (i.e the spread of infection from city to city due to people flying 
long distances by plane). It is thus very important to address the problem
of the mobility of humans and incorporate it into the models to be able to
make more reliable predictions, and then, to be able to propose effective
public policies against the dispersal of a known or emerging
disease.\bigskip 

Including the mobility of the human population in a model is not an easy
task given the complexity of  human behavior. The first problem to address
is the technical and ethical difficulties that arise when trying to get
information about the mobility of humans. There are many databases from
which this data could be inferred, such as the ones associated to cellular
phone networks, credit cards, hotel reservations, flight reservation
databases, etc. But as almost all of them are private, most researchers do
not have access to them. Moreover, even if we did have them, mixing this
diversified information together to get a human mobility model is a hard
task by itself. Aside from this particular difficulties, there is an
intrinsic bias on the databases if we are going to use them for diseases
spread, because its reasonable to think that human behavior will change, or
adapt in presence of social awareness of a disease \cite%
{Epst08,Gros08,Gros06,Risa09,Zane07,Zane07-2,Wang07,Feff07,Funk10,Zhao10},
and the inferences made on this databases can not take that into account.
Moreover there can be a social bias because not everyone may use credit cards, go to hotels, etc. 

Whether it is necessary to have a detailed information on the
movements of each individual to build up a model, or if it is only needed a
coarse grain statistics of the mobility as a whole is an open question wich
still has to be answered \cite{Lisa09}.

Several works tackle the issue of the correct description of the
human beings mobility, relaying on different methods and databases \cite{injo07,bara08,Broc06,Broc07,Chow03,Catt10,Cand07,Gonz09}.

As most works focusing on this topic analyze the effect of the human
mobility in human-human transmitted diseases \cite{Lato08,Funk10,Li110,Zhao10,Feff07,Keel07,Wang07,Zane07-2,Risa09,Gros06},
but not in vector borne ones \cite{Pong08,stod09,Lisa09}, in the present work
we show a variation of our previous dengue model which not only includes the
flight of the mosquitoes but also the mobility of the humans beings. We then
show the key differences in the results between both our models and conclude
on the actual impact of the human behavior on dengue.

In section II we describe the characteristics of the epidemiological model
we use in this work encompassing the dynamics of the virus for humans,
mosquitoes and the dispersal dynamics for each. In section III we present
the results of our numerical investigations which include the analysis of the
size and time evolution of the epidemics, the morphological properties of
the patterns of spatial distributions of the infections for different
mobility patterns and for different densities of mosquitoes. Finally
conclusions are drawn in section IV

\bigskip

\section{The epidemiological model}

There are four ingredients in this model, the epidemiological dynamics of
the infected mosquitoes, the epidemiological dynamics of the infected humans
and the mobility pattern of the individuals and mosquitoes. Each of this
elements will be discussed in what follows

\bigskip

\subsection{Mosquitoes}

The dengue virus does not make any effect to the vector, as such, \emph{%
Aedes aegypti} populations are independent of the presence of the virus. In
the present model mosquitoes populations are produced by the \emph{Aedes
aegypti} model \cite{oter08} with spatial resolution of one block using
climatic data tuned to Buenos Aires, a temperate city where dengue
circulated in the summer season 2008-2009 \cite{seij09}. The urbanistic unit
of the city is the block, approximately a square of (100m x 100m). Because
of the temperate climate the houses are not open as it is often the case in
tropical areas. Mosquitoes develop in the center of the block which often
presents vegetation and communicates the houses of the block. The model then
assumes that mosquitoes belong to the block and not to the houses and they
blood-feed with equal probability in any human resident in the block. \emph{%
Aedes aegypti} is assumed to disperse seeking for places to lay eggs. The
mosquitoes population, number of bites per day, dispersal flights and adult
mortality information per block is obtained from the mosquitoes model \cite%
{oter08}.

The time step of the model has been fixed at one day. 

The virus enters the mosquito when it bites a viremic human with a
probability $p_{hm}(j)$ depending of the day $j$ in the infectious cycle of
the human bitten. The cycle continues with the reproduction of the virus
within the mosquito (extrinsic period) that lasts $\tau _{m}$ days. After
this reproduction period the mosquito becomes infectious and transmits the
virus when it bites with a probability $p_{mh}$. The mosquito follows a
cycle Susceptible, Exposed, Infected (SEI) and does not recover. Eventually mosquitoes die with a daily
mortality of $0.09$ \cite{oter06}. The adult female mosquitoes population as
produced by the \emph{Aedes aegypti} simulation is then split into
susceptible, $\tau _{m}$ stages of exposed and one infective compartment
according to their interaction with the viremic human population and the
number of days elapsed since acquiring the virus.

The mosquito population of each block is not fixed, but instead mosquitoes
move around in terms of a simple diffusion process.

\bigskip

\subsection{Humans}

The evolution of the disease in one individual human, $h$, evolves as
follows:

$%
\begin{array}{ccc}
Day &  &  \\ 
&  &  \\ 
d=d_{0} & \rightarrow & 
\begin{array}{c}
\text{The virus is transmitted by the bite } \\ 
\text{of an infected mosquito.} \\ 
\text{The human enters the exposed stage} \\ 
\end{array}
\\ 
&  &  \\ 
d=d_{0}+\tau _{E}(h) & \rightarrow & 
\begin{array}{c}
\text{The human }h\text{ enters the infective stage} \\ 
\end{array}
\\ 
&  &  \\ 
d=d_{0}+\tau _{E}(h)+j & \rightarrow & 
\begin{array}{c}
\text{The human is infective and } \\ 
\text{transmits the virus to a biting mosquito } \\ 
\text{with probability }p_{hm}(j) \\ 
\end{array}
\\ 
&  &  \\ 
d>d_{0}+\tau _{E}(h)+\tau _{i} & \rightarrow & 
\begin{array}{c}
\text{The human enters the recovered stage. } \\ 
\text{No longer transmits dengue }%
\end{array}%
\end{array}%
$

In the table above $\tau _{E}(h)$ stands for the intrinsic incubation time, and $\tau_{i}$ is the viremic 
time of each individual. 
The cycle in the human being is then of the form Susceptible, Exposed,
Infected, Recovered (SEIR). Each human has it's own value of $\tau _{E}$ 
which is assigned according to the
Nishiura's experimental distribution \cite{nish07}.

As mentioned above, our analysis of the time evolution of dengue fever is
performed on a schematic city in which the basic unit is the block and in
each block a human population of 100 individuals is placed.
An exposed human(index case)
is introduced near the center of the grid, on January $1^{st}$.

The human population of each block is not fixed in the present work. In
order to describe the patterns of mobility of the humans we have adopted the
following schematic model. 50\% of the population of each block is randomly
selected to be mobile, while the other 50\% is considered to remain in its
original block during the whole analysis. Each mobile individual is assumed
to stay $2/3$ of the day in its original block, while the other $1/3$ of the
day she/he will stay in a randomly assigned block according to the
corresponding distribution. Each of the mobile individuals is assigned in
each case a fixed destination to which it returns everyday. At the end of
the day individuals return to their original block.  This random assignment
is performed according to certain rules that will characterize the mobility
pattern. Following recent works on human mobility, referred in the
introduction, we require that the movement of each individual a) should be
highly predictable \cite{Chao10} and b) the distribution of the lengths of
the displacements of the human should follow a truncated Levy
distribution \cite{bara08} which reads:%
\begin{equation}
P(r)\propto(r+r_{0})^{-\beta }\exp (-r/\kappa ).  \label{eq:levy-dist}
\end{equation}
Being $P(r)$ the probability of a human traveling a distance, where $r_0$, $\beta$ and $\kappa$ are parameters
that characterize the distribution.
In this work we have used the parameters described in Table \ref{tab:parameters}.

\begin{table}[tbh]
\caption{Levy-Flight distribution parameters.}
\label{tab:parameters}%
\begin{ruledtabular}
\begin{tabular}{lccr}
 &$r_{0}(m)$ &$\beta$ &$\kappa$ (m)\\
\hline
1 & 100 & 1.65 & 1500 \\ 
2 & 100 & 2. & 1500 \\ 
3 & 100 & 3. & 1500 \\ 
4 & 100 & 4. & 1500 \\ 
\end{tabular}
\end{ruledtabular}
\end{table}

\begin{figure}[htb]
\includegraphics[scale=0.4, angle=-90]{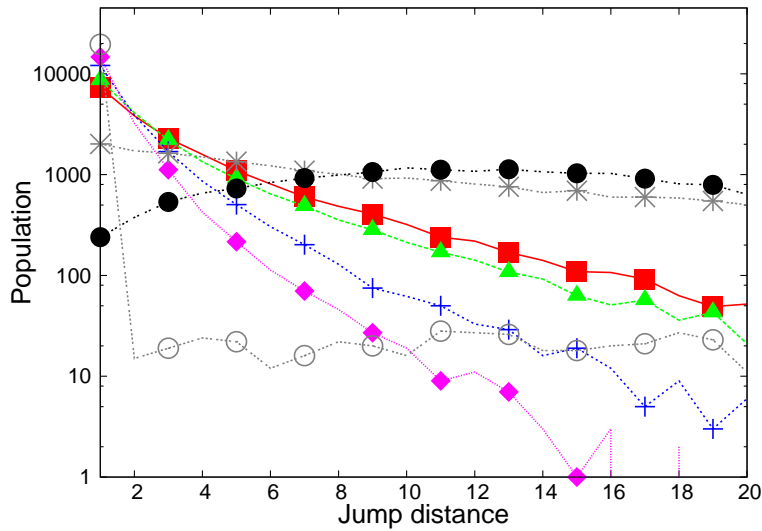} 
\caption{Distribution of jump length for
different mobility patterns. Random(Full circles), Move400(Empty circles), 
Levy-Flights with $\beta=1.65$(Full squares), $\beta=2$(Full triangles), $\beta=3$(Crosses), $\beta=4$(Full diamonds).
As comparison, with asterisk is shown the Levy-Flight with parameters from the Barabasi distribution \cite{bara08}.}
\label{fig:distribuciones}
\end{figure}

Such a pattern of mobility of the humans is accomplished by building a
network with 50 links starting in each block. The length of the link is
distributed according to the proposed length distribution and the final
block is chosen at random from those which can be reached by the link. Each
link is assigned to a mobile human at the start of the simulation. The
distribution of jumps sizes for each type of underlying network is shown on
Fig \ref{fig:distribuciones}.

In order to have reference mobility networks we have also analyzed the case
in which the endpoints of the links are completely random. We have also
investigated the case in which only one individual per block performs a
random jump while the rest of the mobile individuals in the block visit only
their neighboring blocks ($Move400$).

Once the set of parameters is fixed, different underlying networks
are generated and a set of evolutions (typically a couple of thousand
events) is performed for such arrangements.

\bigskip

\subsection{The networks}

The networks built according to the above mentioned prescription can be
analyzed in order to unveil their "small world properties". We have found it
interesting to study the geodesic path i.e. the average minimum path between
all the cells.%
\begin{equation}
l=\frac{1}{n(n-1)}\sum_{i,j\neq i}d_{ij}
\end{equation}

with $d_{ij}$ the minimum path between cells $i$ and $j$, and $n$ the total number of nodes.
The minimum
path is defined as the minimum number of links that are to be traversed in
order to travel from the original block to the destiny block.Therefore, we see that
it is a simple average over all the possible pairs of blocks in the system
of the minimum distance between each pair.

\bigskip

Another interesting magnitude to explore the characteristics of a network is
the so called Clusterization. One of the usual definitions of this magnitude
is:

Given a node $i$  with $k_{i}$ nearest neighbors we define $c_{i}$ as 
\begin{equation}
c_{i}=\frac{2}{k_{i}(k_{i}-1)}\left( \text{number of links between nearest neighbors}%
\right) 
\end{equation}

\begin{equation}
C=\frac{1}{n}\sum_{i}c_{i}
\end{equation}

\bigskip

In Table \ref{tab:cluster} we show the results of such a calculation.  

\begin{table}[htb]
\caption{Clusterization and Mean shortest path for
the human mobility networks.}
\label{tab:cluster}%
\begin{ruledtabular}
\begin{tabular}{lcr}
$Network$ &$l$&$C$\\
\hline
Random & 1.779 & 0.220\\
Levy($\beta=1.65$) & 2.006 & 0.286\\
Levy($\beta=2$) & 2.112 & 0.303\\
Levy($\beta=3$) & 2.532 & 0.349\\
Levy($\beta=4$) & 3.092 & 0.380\\
Only $400$ move & 3.884 & 0.014\\
\end{tabular}
\end{ruledtabular}
\end{table}

We can see from table I that the mean shortest path attains a
minimum for the completely random pattern of links and grows as this pattern
is replaced by the ones generated by the levy flights. We see that
the broader the Levy-Flight is the larger is the average minimum path, as expected.

\bigskip 

\section{Numerical calculations}

We have implemented the above described method (The algorithm without human
mobility has been fully described in \cite{oter11}) and have performed extensive
calculations for different initial conditions.

The conditions are : mosquitoes breeding sites density, different
realizations of the underlying mobility networks and different seasonal
conditions.

The number of breeding sites per block explored in this calculations are 50,
100,200,300 and 400. Larger number of breeding sites are considered to be
too unrealistic for the system we have in mind i.e. the city of Buenos
Aires. Moreover larger number of breeding sites do not add new information to
our calculations.

Another relevant condition that we have explored is the underlying mobility
network. For each value of the breeding sites density, evolutions with
different underlying networks were performed.

\begin{figure}[hbt]
\includegraphics[width=12cm,angle=-90]{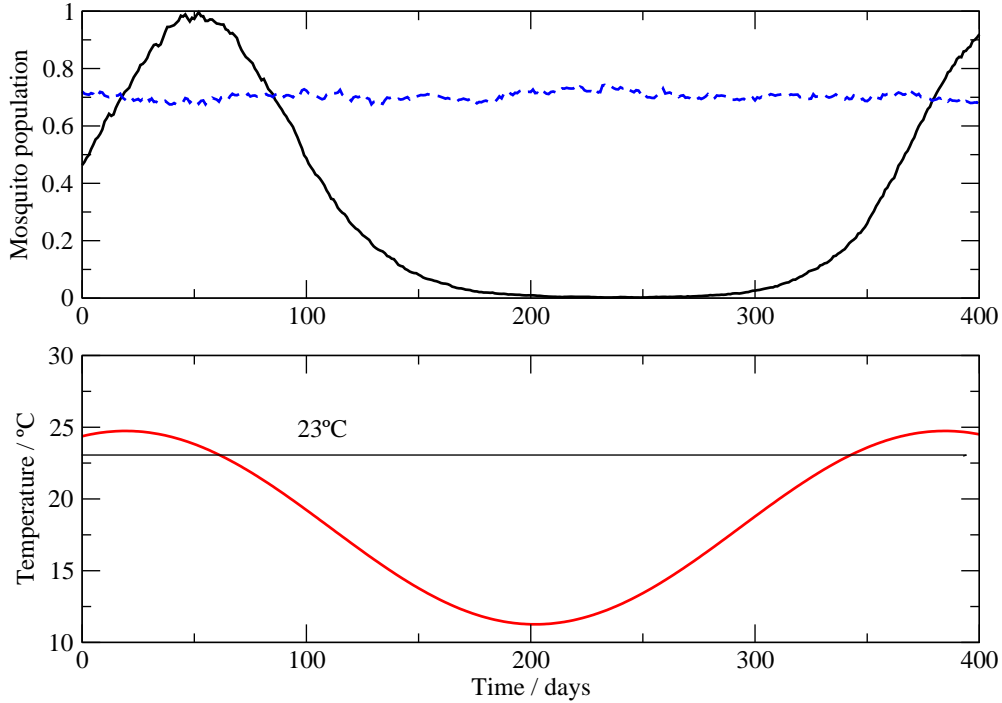}
\caption{Top: Normalized mosquitoes population for constant temperature(Dashed) and seasonal variation(full). 
Bottom: Temperature profile for constant and seasonal variation. Day $0$ is set at $1^{st}$ of January.}
\label{tempymosq}
\end{figure}

Finally we have considered two different seasonal conditions, on Fig. \ref{tempymosq} we show the population(top)
and temperature(bottom) profiles. 
In the season-less situation, the
temperature remains fixed all along the evolution at $23^{0}C$. In this case
the mosquito population remains basically constant all along the
evolution. In such a case the size of the epidemics is determined by the
dynamics of the infection itself subject to the above mentioned boundary
conditions. If we adopt the average temperature time-distribution of BA (see
figure for details) the population of mosquitoes is a strongly time
dependent one, the size of the epidemics might then be severely constrained.

\bigskip

\bigskip

In what follows we will focus on certain properties of the epidemic system that are relevant for the 
understanding of the characteristics of the time evolution. In first place we study the morphology
of the evolving spatial structure of the epidemics. Then we study the final size and time span of the epidemics. 
Then we include the results of the analysis of a system in which the temperature is kept
fixed at $23$ degrees Celsius. Finally we study a new magnitude that we name the $power$ of the 
epidemics. 

\bigskip

\bigskip

\subsection{Morphology of the spatial structure of the epidemics}

\bigskip
It is expected that human mobility increases
the size and speed of the epidemics.

This happens because each jump (shortcut) when executed by a
virus carrying individual may induce the contagion of mosquitoes at the
destination block and then generates a new dispersal center for the illness.

\begin{figure}[htb]
\includegraphics[width=0.3\linewidth, angle=-90]{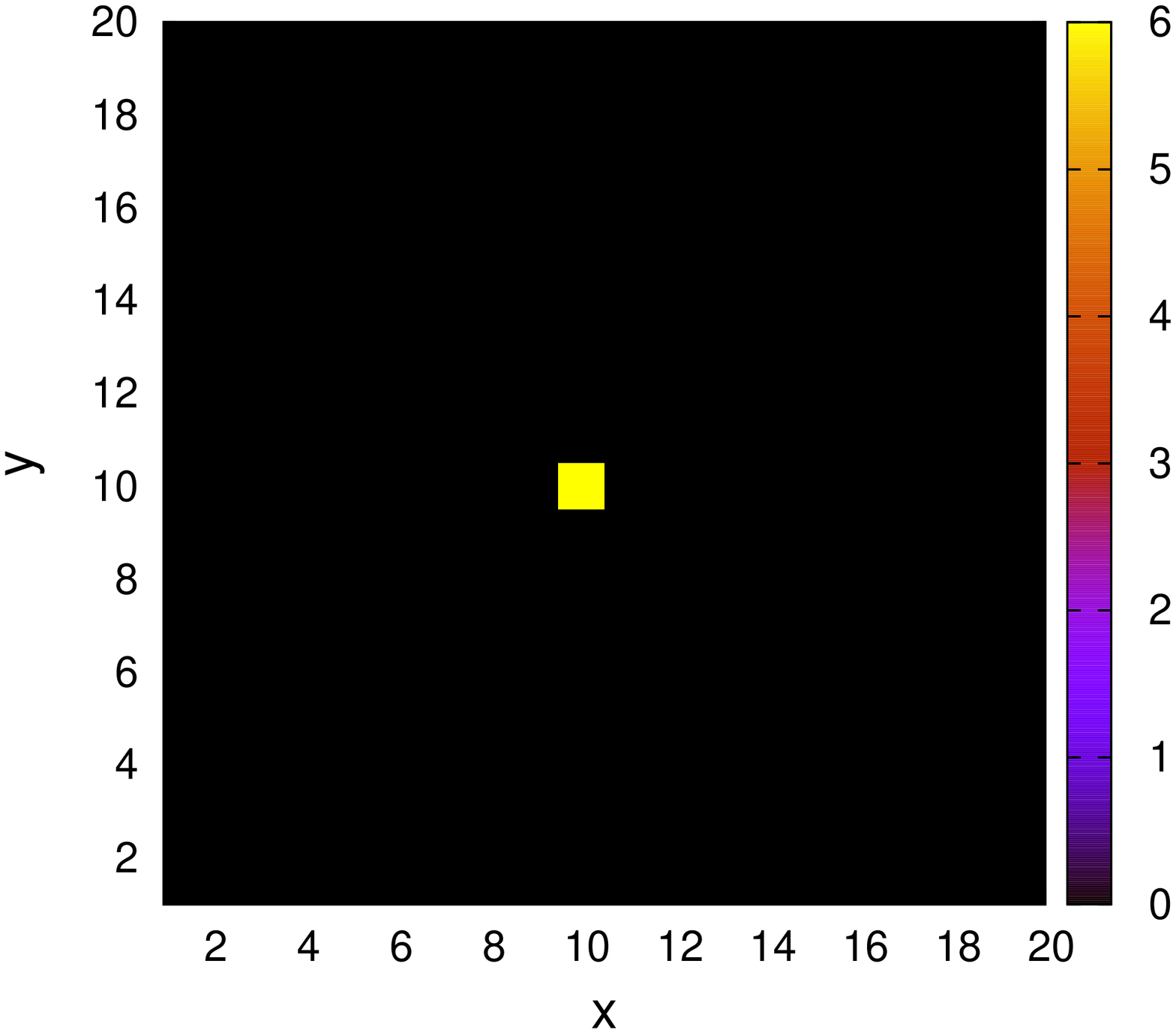} %
\includegraphics[width=0.3\linewidth, angle=-90]{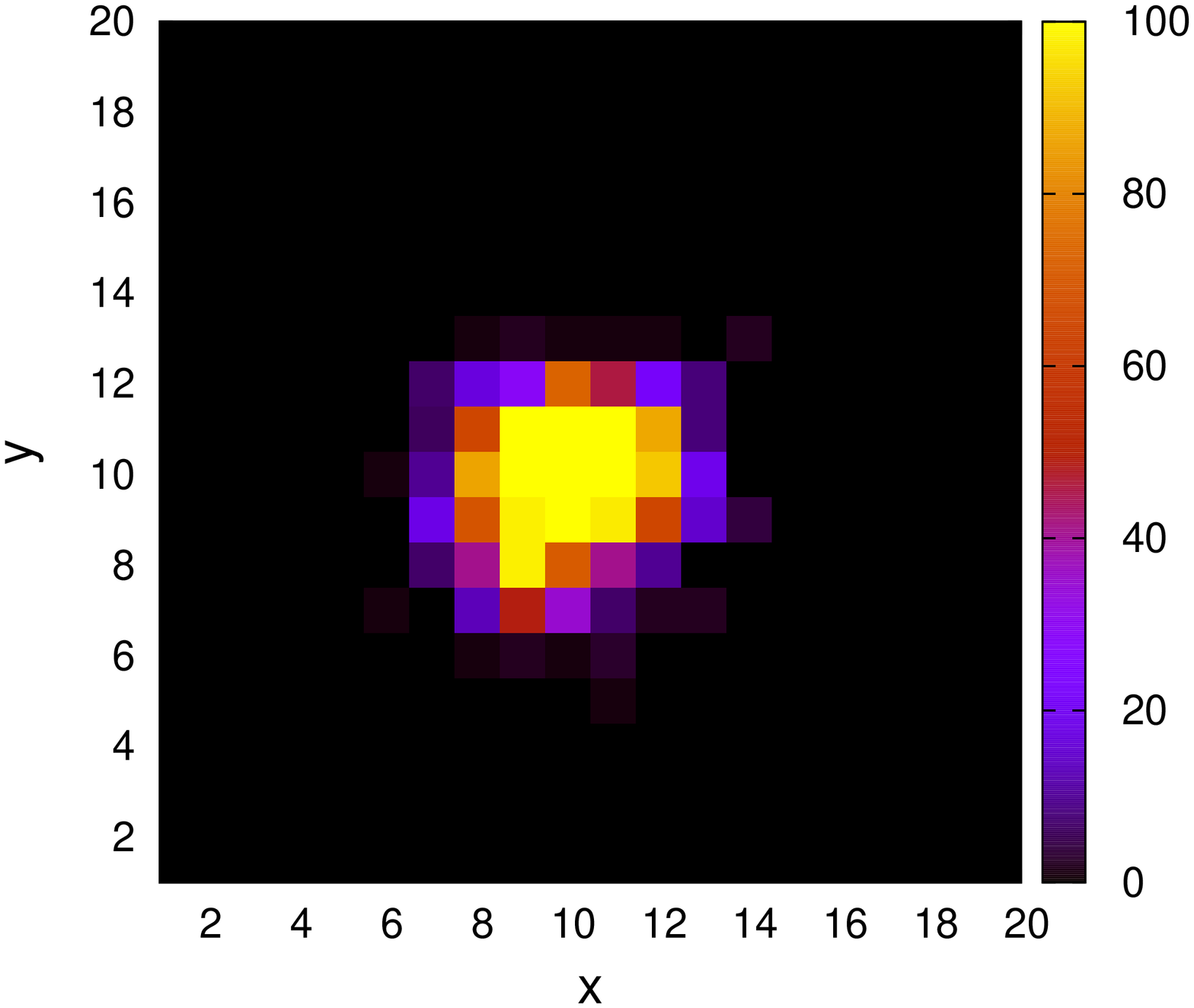} %
\includegraphics[width=0.3\linewidth, angle=-90]{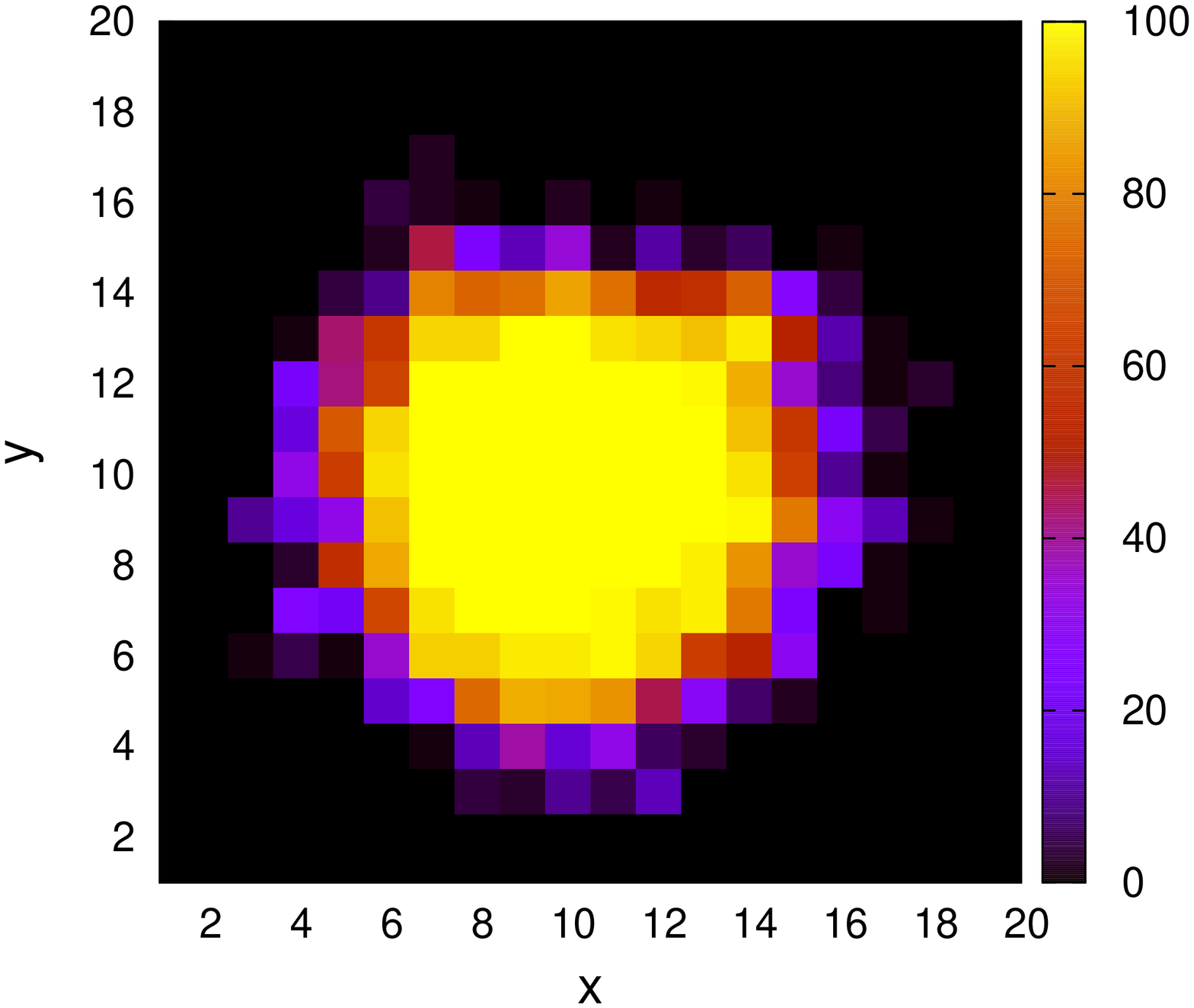}  
\caption{
Spatial distribution of the number of recovered individuals for three 
times namely 25 days, 81 days and 173 days for the case of a dengue epidemic driven only by the 
dispersal of mosquitoes. It can be seen that the pattern corresponds to a diffusive process and
the evolution is highly symmetric.}
\label{fig:diff-secuencia}
\end{figure}

In Fig \ref{fig:diff-secuencia} we show the density of recovered individuals
at three relevant times for the case in which the dispersal of dengue is
driven by the diffusion of mosquitoes only. It is seen that the population of
recovered individuals displays a symmetrical pattern as expected from a
simple diffusion process.

\bigskip

\begin{figure}[htb]
\includegraphics[width=0.3\linewidth, angle=-90]{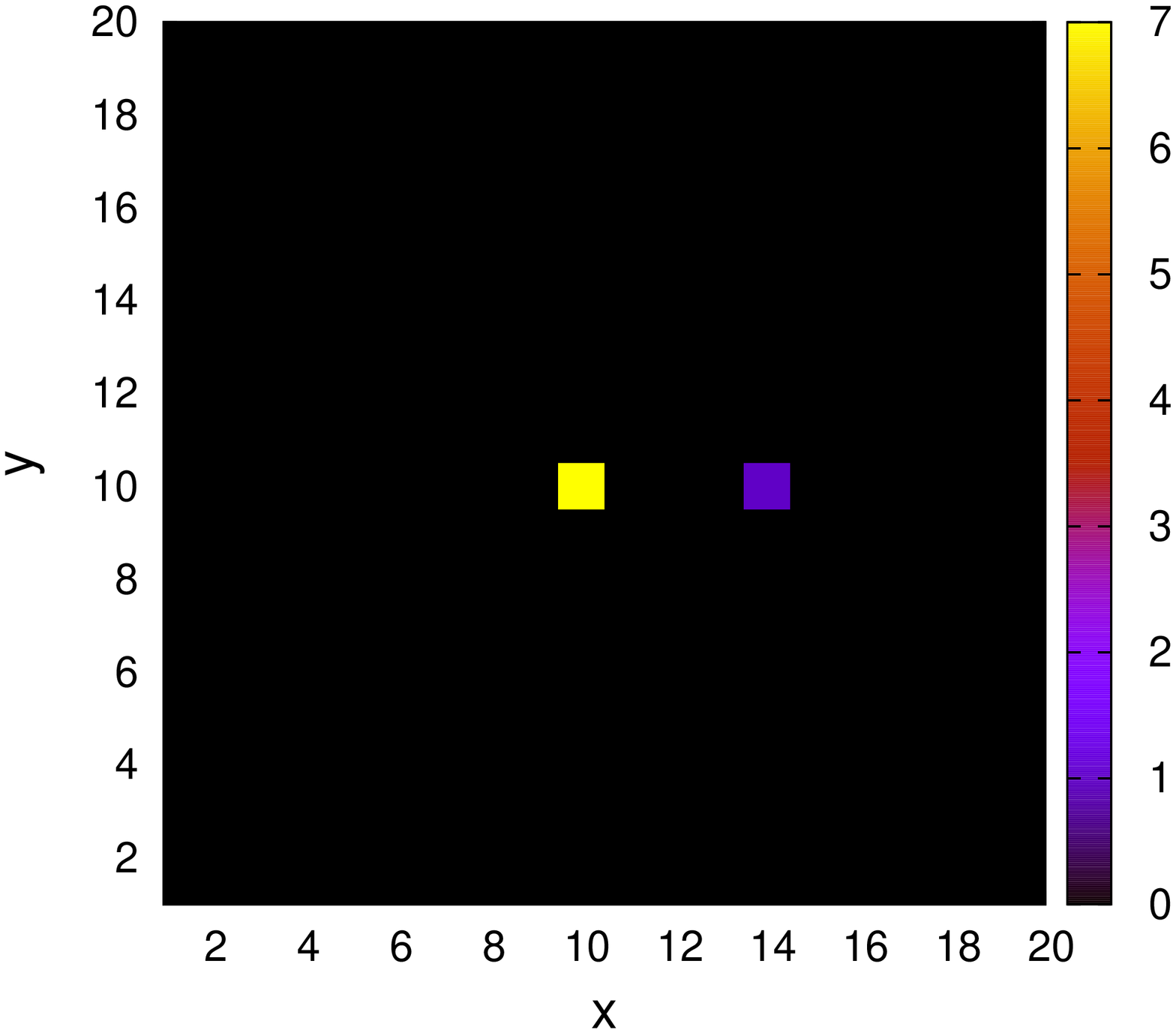} %
\includegraphics[width=0.3\linewidth, angle=-90]{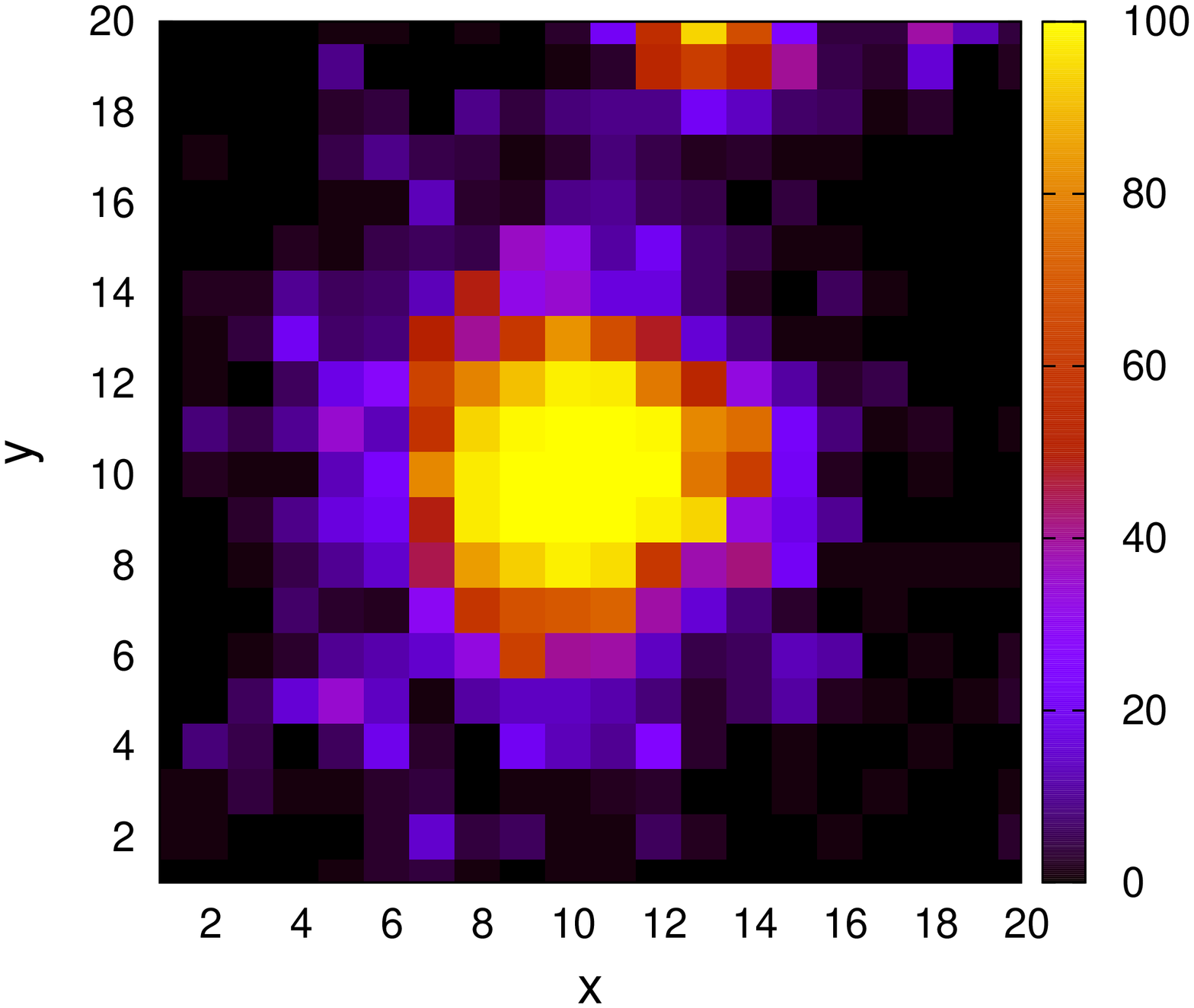} %
\includegraphics[width=0.3\linewidth, angle=-90]{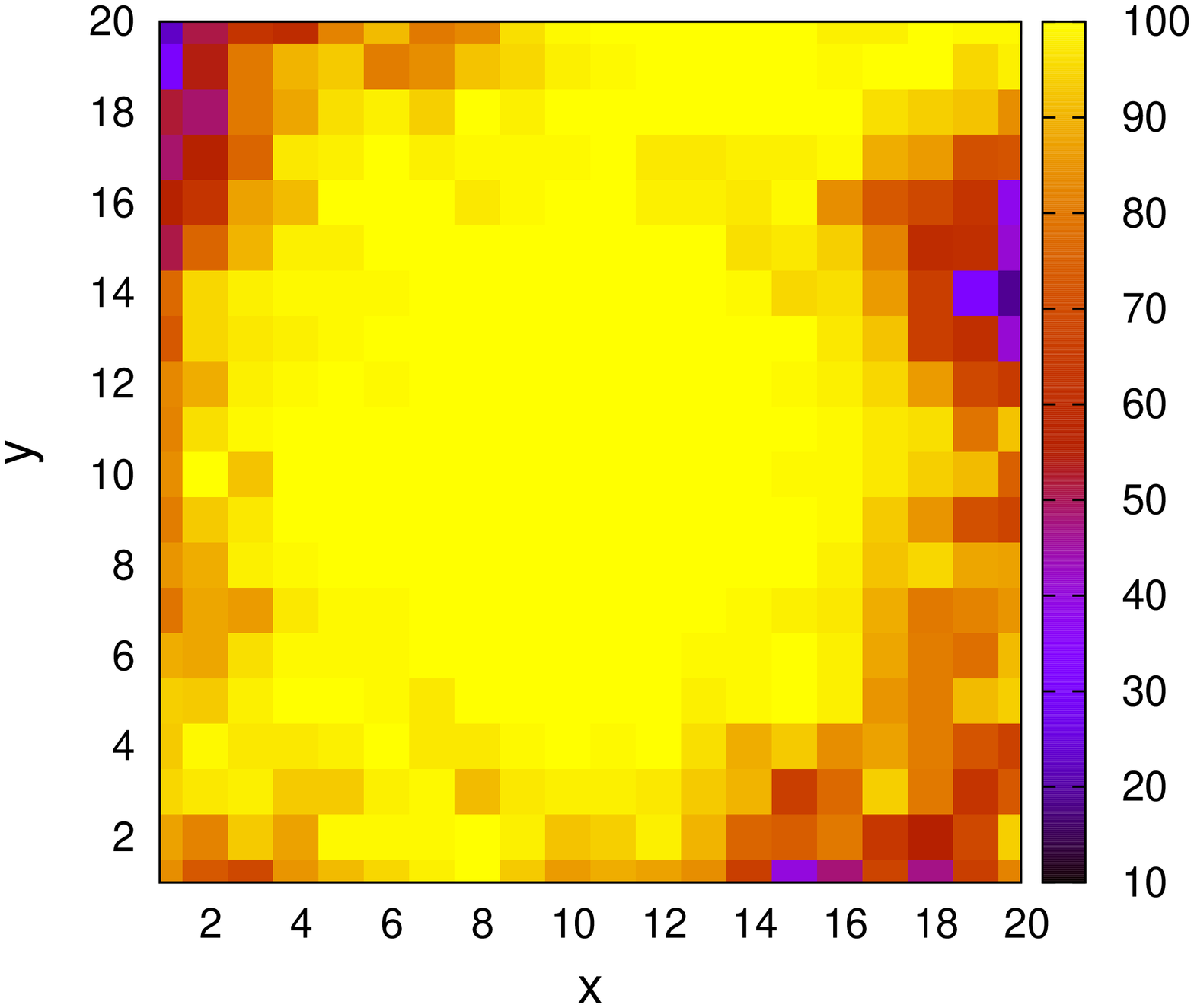}  
\caption{
For the same times as in Fig. \ref{fig:diff-secuencia}, the spatial distribution of recovered 
individuals when their mobility is modeled by a Truncated Levy-Flight with $(\beta)=3$. It can be 
seen that, at variance with Fig. \ref{fig:diff-secuencia} the pattern is asymmetrical, moreover the evolution is
faster and it can be seen that at $t=81$ days a second focus is present. 
}
\label{fig:levy3-secuencia}
\end{figure}

On the contrary as seen in Fig \ref{fig:levy3-secuencia} the pattern for the
case in which the human jumps follow a Levy-Flight distribution is quite
heterogeneous and it can be clearly seen that there is more than one
dispersal center.

\bigskip

This observations can be made more quantitative if we calculate the radial
correlation function defined as the probability of finding at least one
infected (recovered) individual (calculated at the time at which all individuals
have returned home) at a block such that it can be reached by a jump 
of length $r$ from the place at which the initial infected individual one was located
(which in this case is a block close to the center of the city).

In what follows we show the result of calculating the radial correlation function for three typical 
cases namely for the mosquito driven evolution, for the case Levy-Flight with $\beta=3$ and the case
in which only one of the mobile individuals performs random jumps.

\bigskip

\begin{figure}[htb]
\includegraphics[scale=0.4, angle=-90]{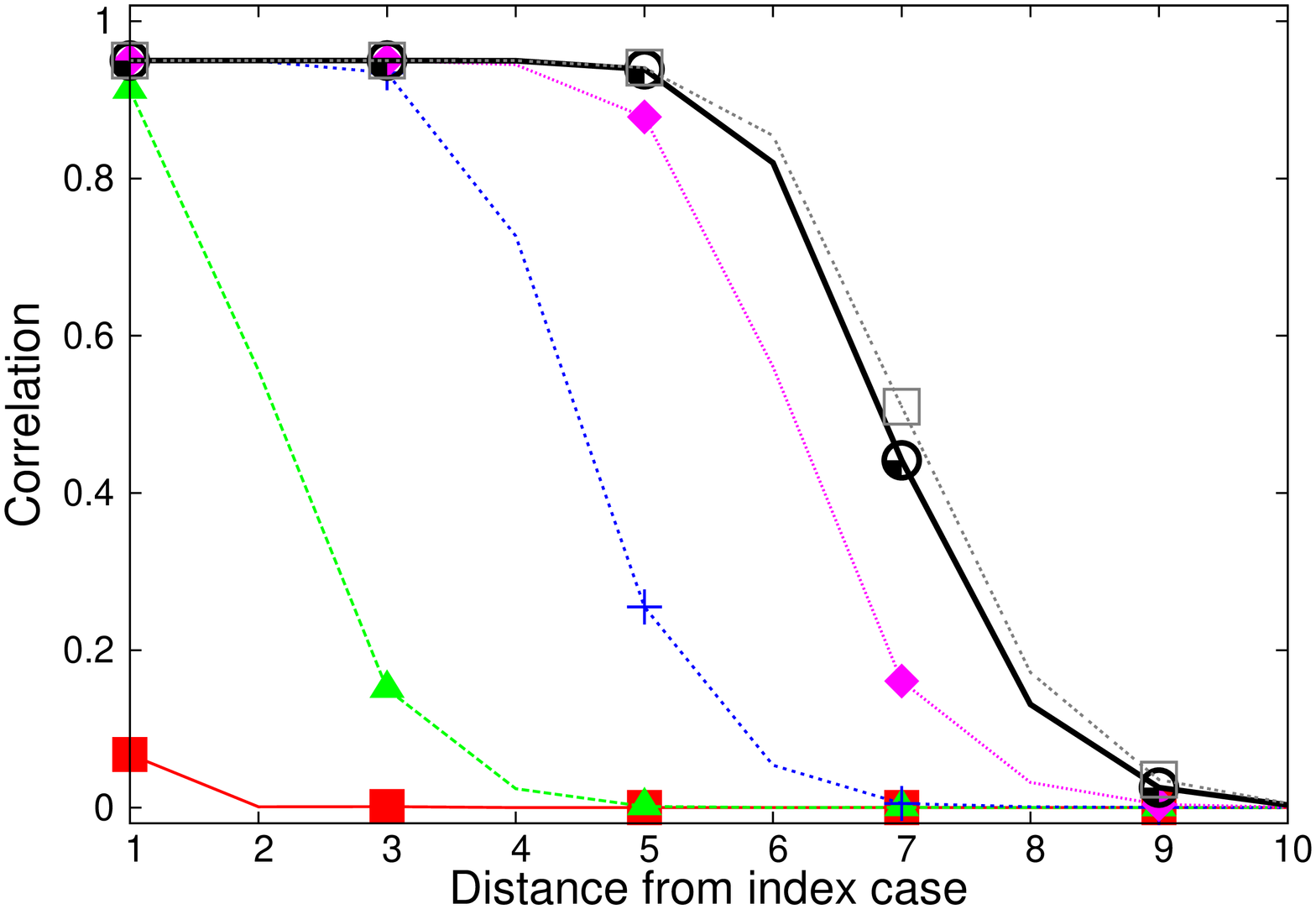} 
\caption{
Radial correlation function as defined in the text for the case in which
the spatial evolution of the epidemics is driven by the dispersal of mosquitoes only. Full squares 
(red)$t=25days$, full triangles (green)$t=57days$, crosses (blue)$t=89days$, diamond (violet)$t=121days$ open circles(black)$t=153days$, open squares (gray) $t=189days$.
The pattern of variation is quite regular as expected for a diffusive case.
}
\label{fig:corr-diff}
\end{figure}

\begin{figure}[htb]
\includegraphics[scale=0.4, angle=-90]{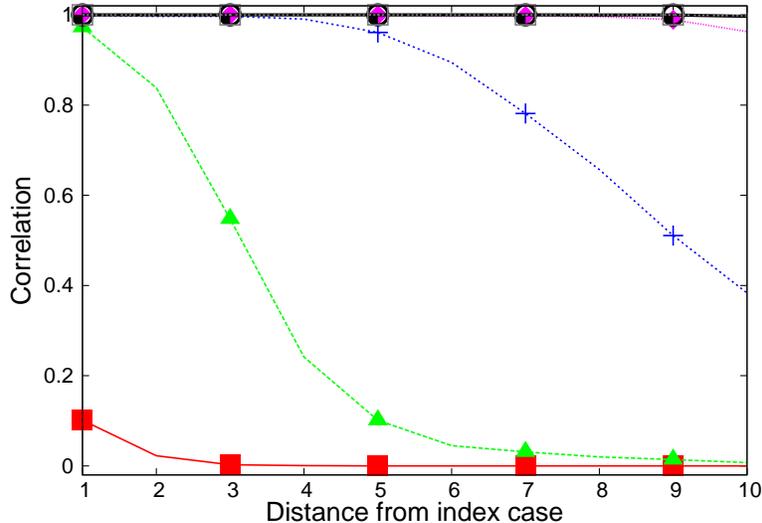} 
\caption{
Radial correlation function as defined in the text for the case in which
the humans move with the Levy-Flight($\beta=3$) distribution. Full squares 
(red)$t=25days$, full triangles (green)$t=57days$, crosses (blue)$t=89days$, diamond (violet)$t=121days$ open circles(black)$t=153days$, open squares (gray) $t=189days$.
}
\label{fig:corr-levy3}
\end{figure}

\begin{figure}[htb]
\includegraphics[scale=0.4, angle=-90]{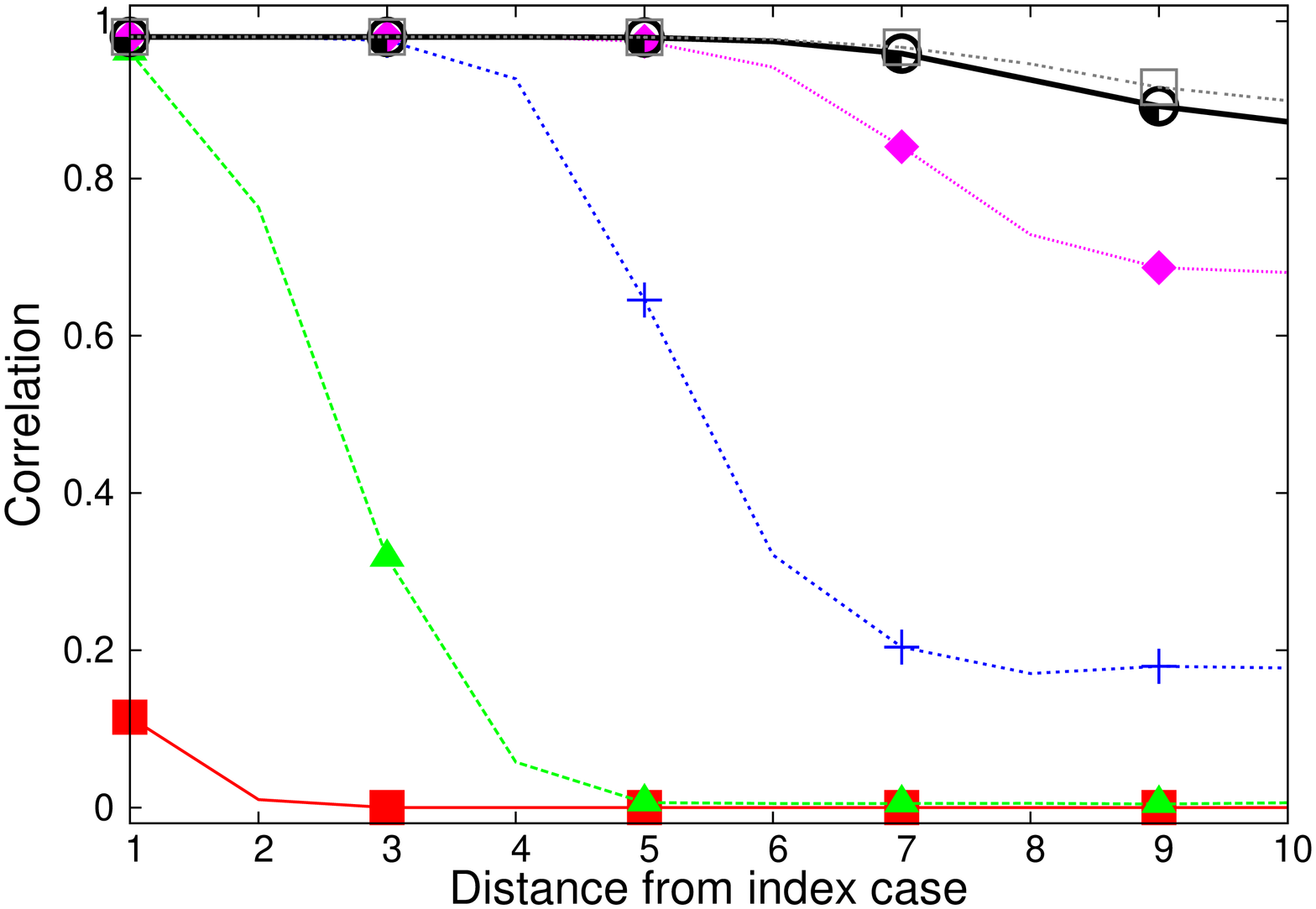} 
\caption{
Radial correlation function as defined in the text for the case in which
only one human per block moves at random, and the rest move to neighboring blocks. Full squares 
(red)$t=25days$, full triangles (green)$t=57days$, crosses (blue)$t=89days$, diamond (violet)$t=121days$ open circles(black)$t=153days$, open squares (gray) $t=189days$.
}
\label{fig:corr-muev400}
\end{figure}

From Figs. \ref{fig:corr-diff},\ref{fig:corr-levy3},\ref{fig:corr-muev400} is
clearly seen that in the presence of human mobility nearly all of the city can be reached by the 
epidemic in short times. On the one hand the case of mosquitoes only as driving force the 
correlation function displays patterns expected for a traveling wavefront. In the other cases the 
wavefront breaks early in the evolution and the correlation function is different from zero almost 
everywhere after a few days.

\bigskip

As we have seen in Fig.3) and 4) the structure of the spatial density of, say, humans in state R is 
highly symmetric and compact for the case without human mobility. As human mobility 
(of the kind considered in this work) is incorporated both the symmetry and the compactness are
lost. In order to explore this behavior in a more quantitative way we define cluster of recovered 
individuals in the following way. Given a block $i$ we will call it an occupied block if at least one 
member of its original population is in the recovered state. A cluster (of size
larger that one) is a set of occupied blocks in
which all constituents have at least a nearest or second nearest neighbor which belongs to the cluster. 
Then the block $i$ will 
belong to the cluster if the following relation is satisfied: 

 
\begin{equation}
i\in C\Longleftrightarrow \exists \text{ }j\in C\text{ }/\text{ }i\text{ is
neighbor of }j
\end{equation}

We define the mass of a cluster as the number of recovered individuals in the cluster.

\bigskip

\begin{figure}[ht]
\includegraphics[width=0.5\linewidth,angle=-90]{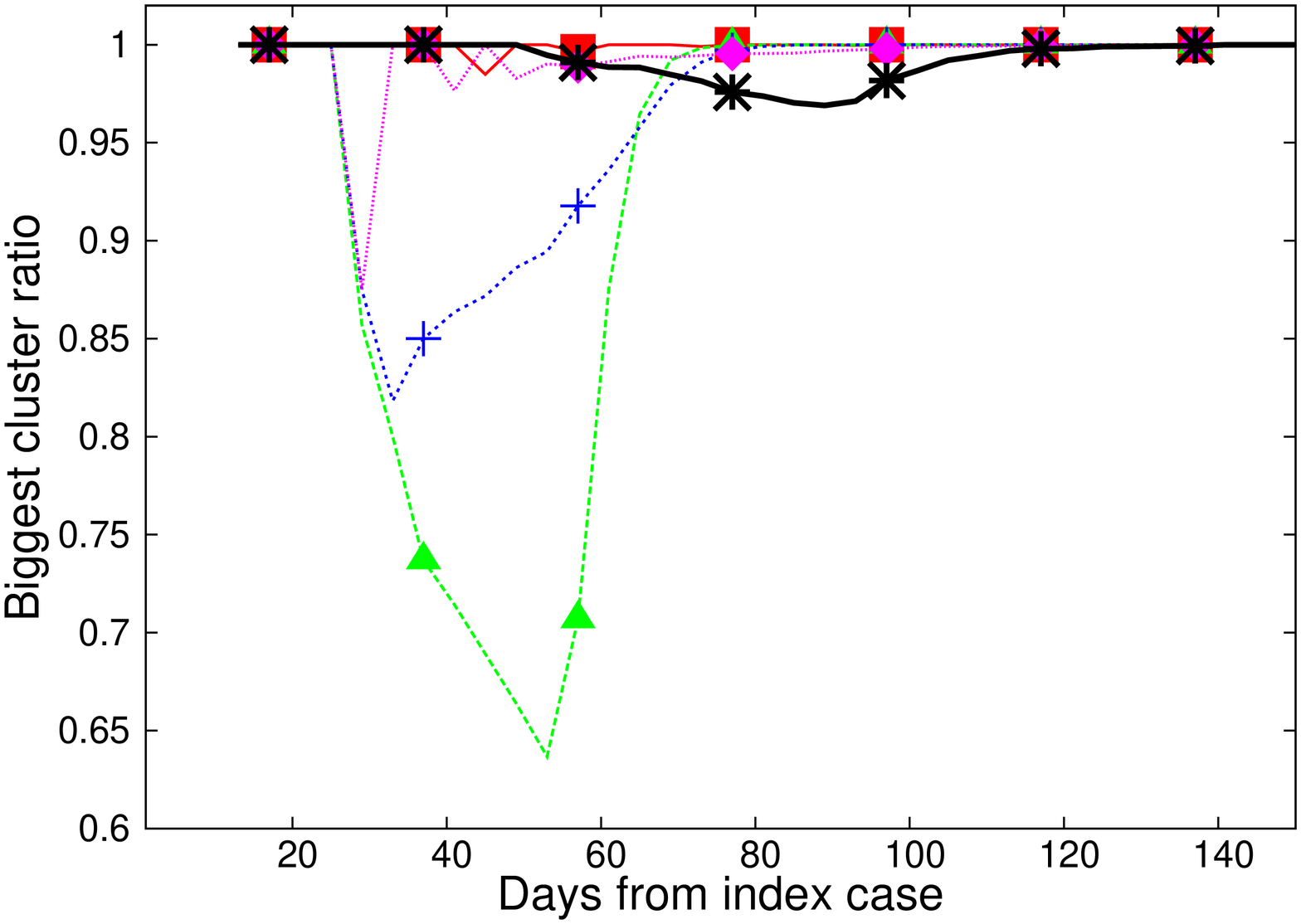}
\caption{
Ratio between the size of the biggest cluster (in terms of the number of 
recovered humans) and the total number of recovered. Diffusive (Red full squares), random (Green full triangles),
 Levy-Flight($\beta=1.65$) (blue crosses), Levy-Flight($\beta=4$) (violet full diamonds), Move400 (Black asterisks).  
}
\label{fig:ratio}
\end{figure}

The results are displayed in the Fig \ref{fig:ratio}.

It can be seen that for the case of simple diffusion all of the mass is
concentrated in the biggest cluster . On the other hand for the
random case there is a time (around $50$ days into the epidemic) at which only about $65\%$ of the 
mass is in the biggest cluster . This is 
due to the emergence of secondary foci generated by infective humans who perform long jumps.

\subsection{Sizes and time span of the epidemics}

\begin{figure}[hbt]
\includegraphics[width=11cm, angle=-90]{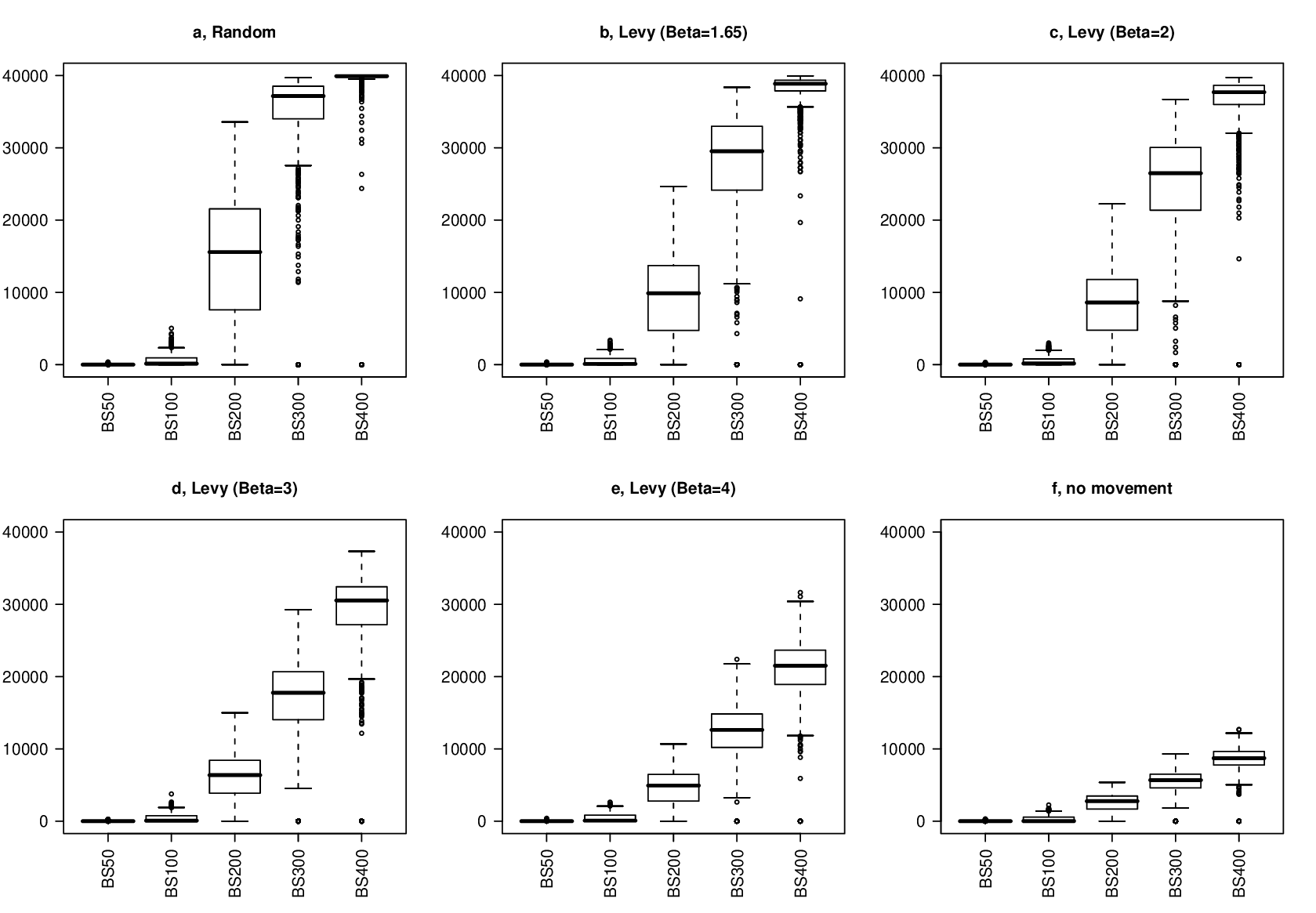}
\caption{Boxplots graphs of the epidemic size for different patterns of
human mobility: (a) random, (b) levy($\beta$=1.65), (c) levy($\beta$=2), (d) levy($\beta$=3), (e) levy($\beta$=4)
and (f) move400.}
\label{marce1}
\end{figure}

One of the main observables in this kind of problems is the final size of the epidemics. In what
follows we show Figure \ref{marce1} a comparison of the final size of the epidemics
in terms of the boxplots corresponding to different values of the breeding
sites density for different patterns of human mobility. As described above we
have two limiting situations $a)$ the case in which the moving humans perform
jumps with completely random destinations and $f)$ the case in
which the dispersal of the epidemics is only due to the diffusion of
mosquitoes. In between we have the patterns related to the
length of the jump given by a truncated Levy-Flight characterized by the
different set of parameters shown in table \ref{tab:parameters} i.e. $\beta =$ $1.65,2,3$ and $%
4$ (keep in mind the smaller $\beta $ the closer to the random
case). The corresponding results are displayed in panels $b-e$. 

\begin{figure}[hbt]
\includegraphics[width=12cm, angle=-90]{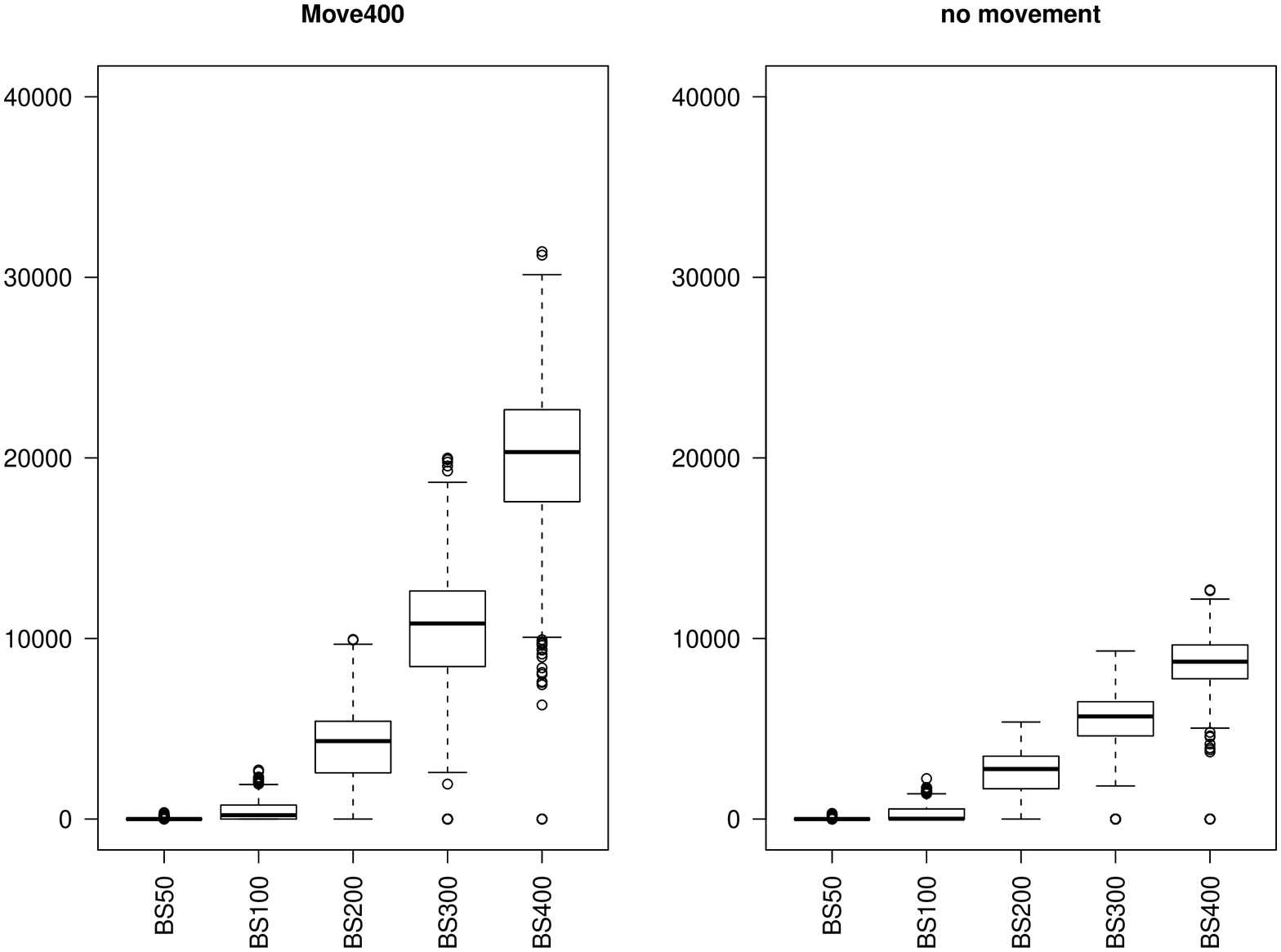}
\caption{Boxplots graphs of the epidemic size for different patterns of
human mobility: (Left) move400 and (Right) no human mobility.}
\label{marce2}
\end{figure}

Finally in Fig. \ref{marce2} we show (left panel)  the case in which only one of the mobile humans in each 
block performs a jump to a random destination while the others move to nearest neighbors. For the sake 
of completeness we show on the right panel the boxplot corresponding to the mosquito only driven evolution.

It is immediate to see that the effect of human mobility for all the cases
is to increase the final size of the epidemics with respect to the case in
which the mosquitoes diffusion is the only driving force. Moreover in the
case of completely random mobility and for the Levy-Flight with $\beta =1.65$
we get that for the highest BS density proposed in this work, the epidemics
spreads over the whole population.

\bigskip



\begin{figure}[hbt]
\includegraphics[width=11cm, angle=-90]{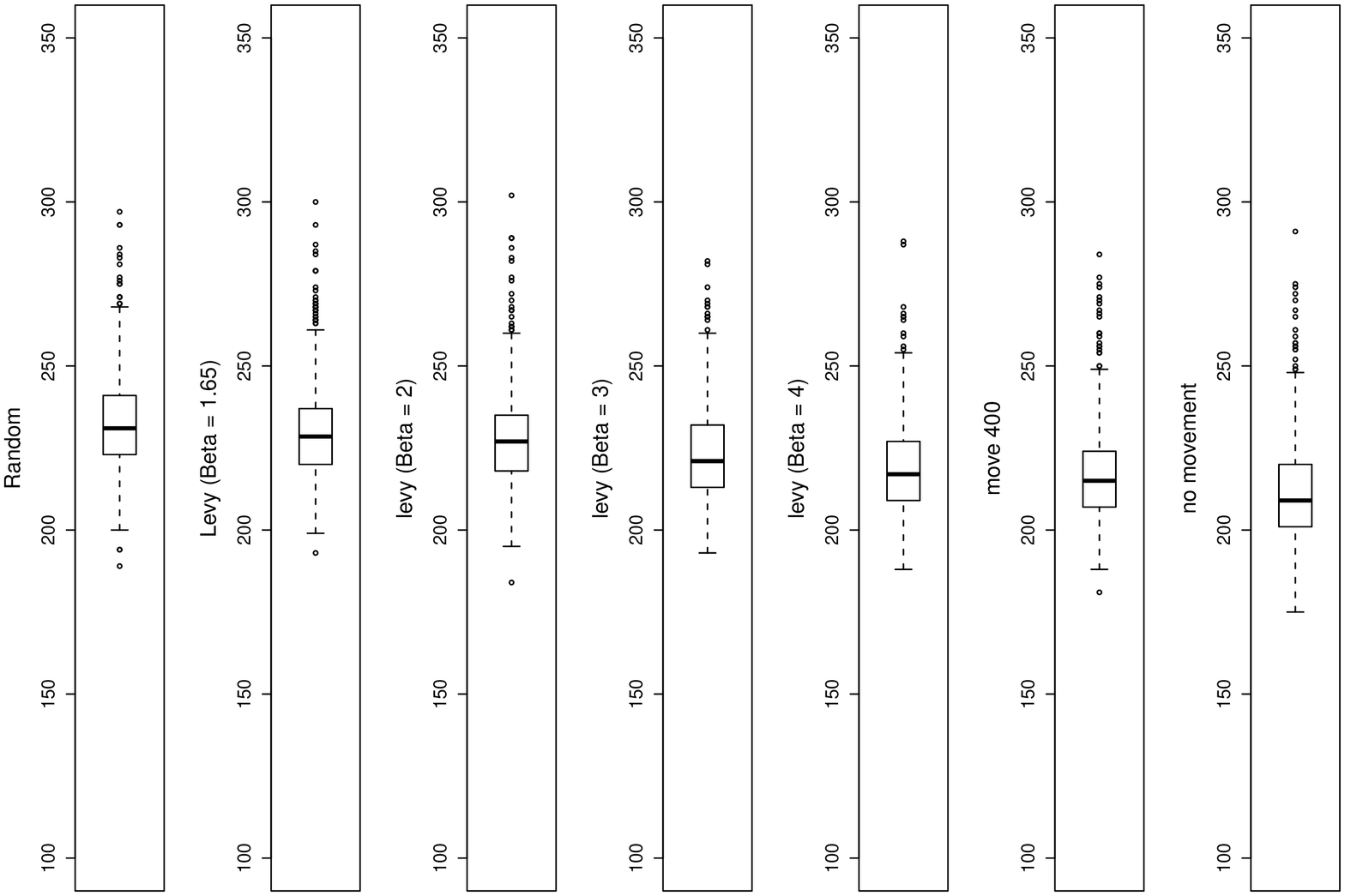}%
\\
\includegraphics[width=11cm,
angle=-90]{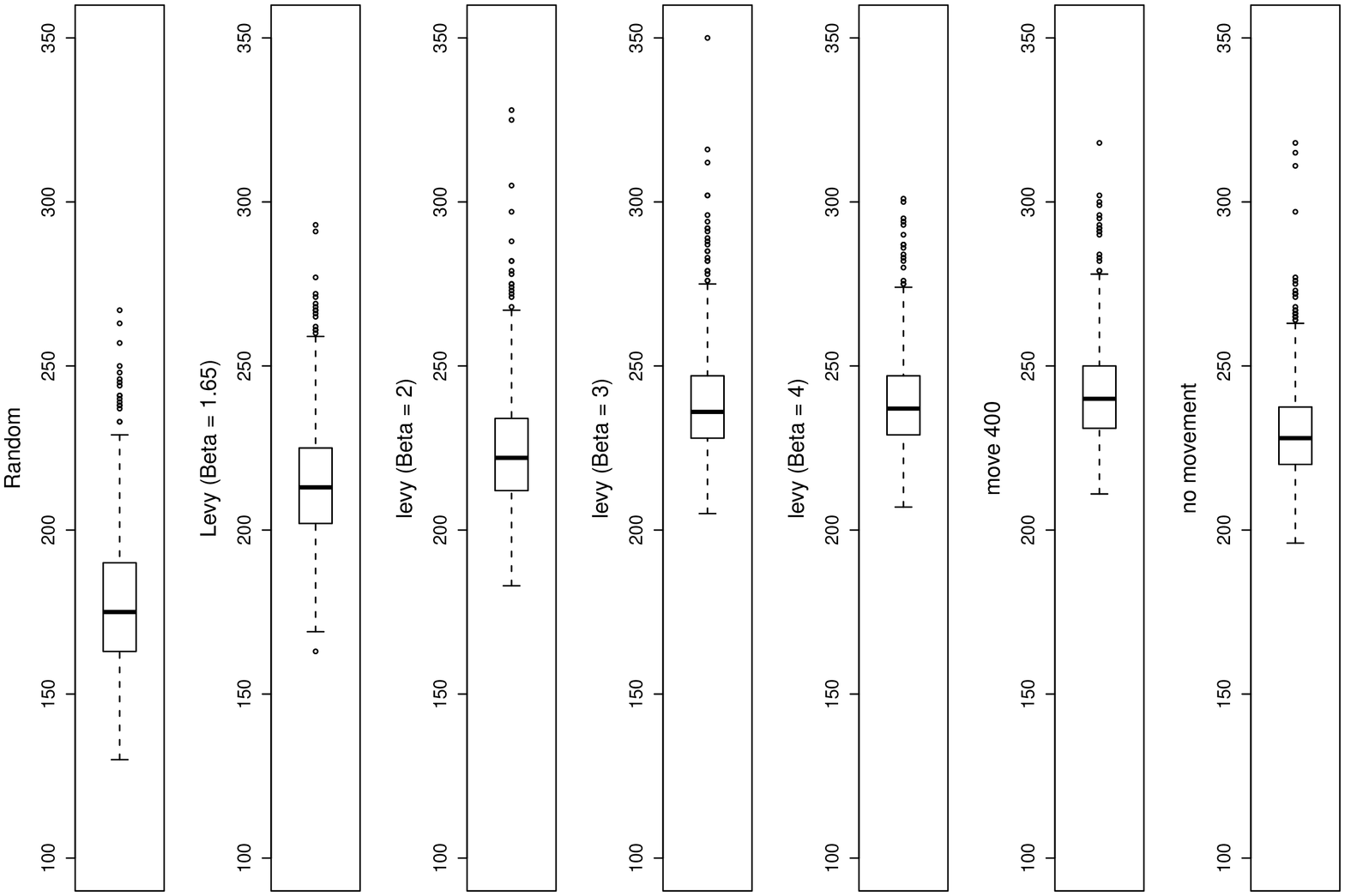}
\caption{Comparison of the duration of the epidemic outbreaks for the
different patterns of human mobility, 200 breeding sites/ha (Top) and 400
breeding sites/ha (Bottom).}
\label{marce3}
\end{figure}

Figure \ref{marce3} shows the duration distribution of the epidemics as a
function of the different patterns of human mobility and for two constant
breeding site densities of 200 BS/ha (Top) and 400 BS/ha (Bottom)
respectively. For 200 $BS$ (Top) all boxplots present a similar spread of
data but a slight tendency of increase of the median from left to right
(from the case without mobility to the complete random pattern). Instead,
for 400 $BS$ the tendency is to decrease from left to right. 
(This behavior will be properly discussed in terms of the Power of the epidemics, see below)

\subsection{Behavior of the model at constant temperature}

\begin{figure}[hbt]
\includegraphics[width=12cm, angle=-90]{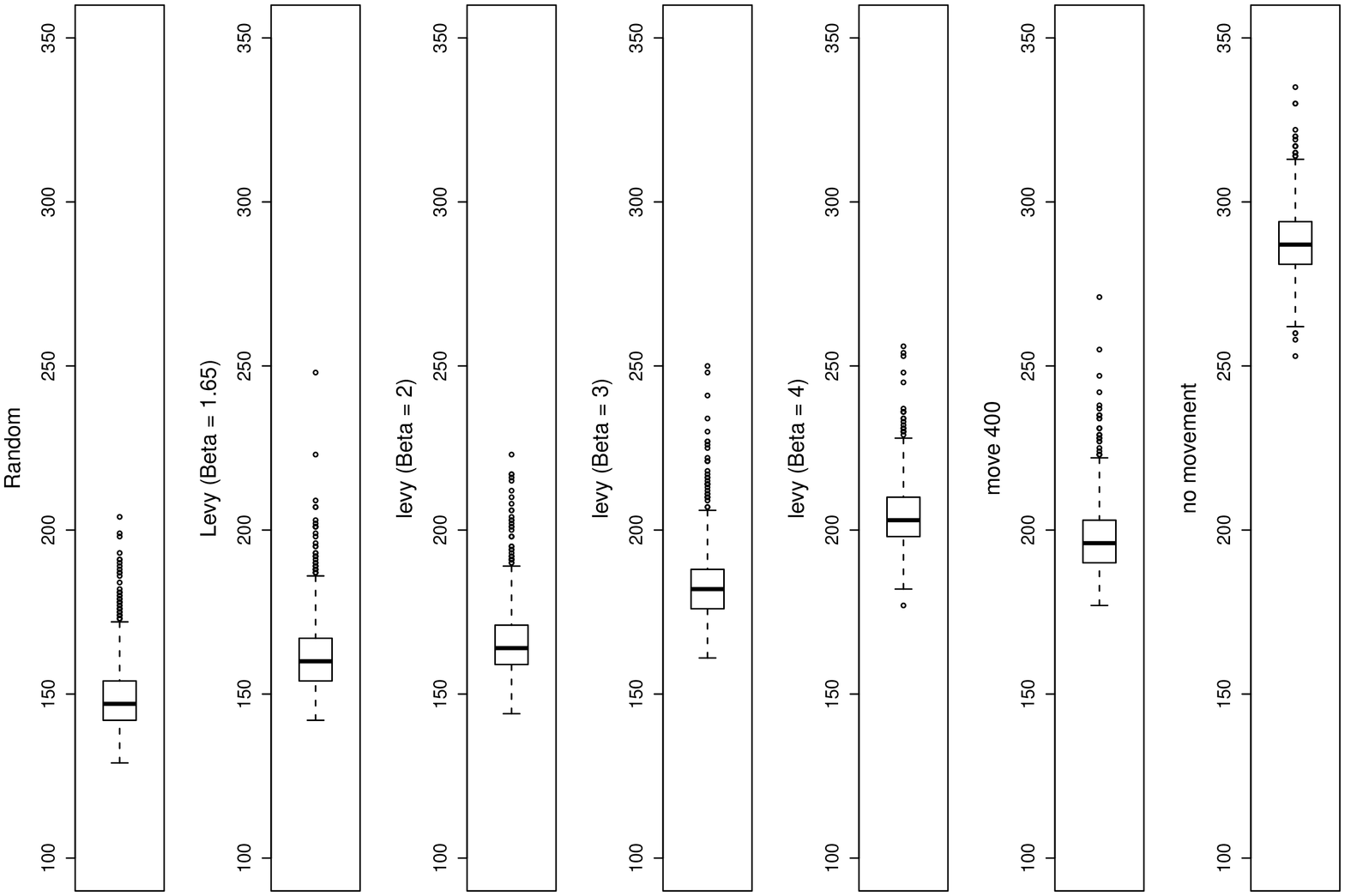}
\caption{Comparison of the duration of epidemics for the different patterns
of human mobility at a constant temperature of 23 degrees Celsius (400
breeding sites / ha).}
\label{marce4}
\end{figure}

Figure \ref{marce4} shows the duration distribution of the epidemics as a
function of the different patterns of human mobility for a constant breeding
site density of 400 BS/ha and a constant temperature of 23 degrees Celsius.
We see that the maximum duration of the epidemic takes place for the case of
no human mobility and it shortens as the patterns of mobility of the
individuals tend to the completely random one. It is also interesting to note that
in this case the epidemics involves the whole population.

\subsection{Power of the epidemic}

\bigskip 

We define the mean power of the epidemic as the ratio between the median of
the final size of the epidemic and the median of the duration of the epidemics.

\begin{equation}
P_{m}=\frac{S}{\tau }
\end{equation}

\begin{figure}[hbt]
\includegraphics[width=12cm, angle=-90]{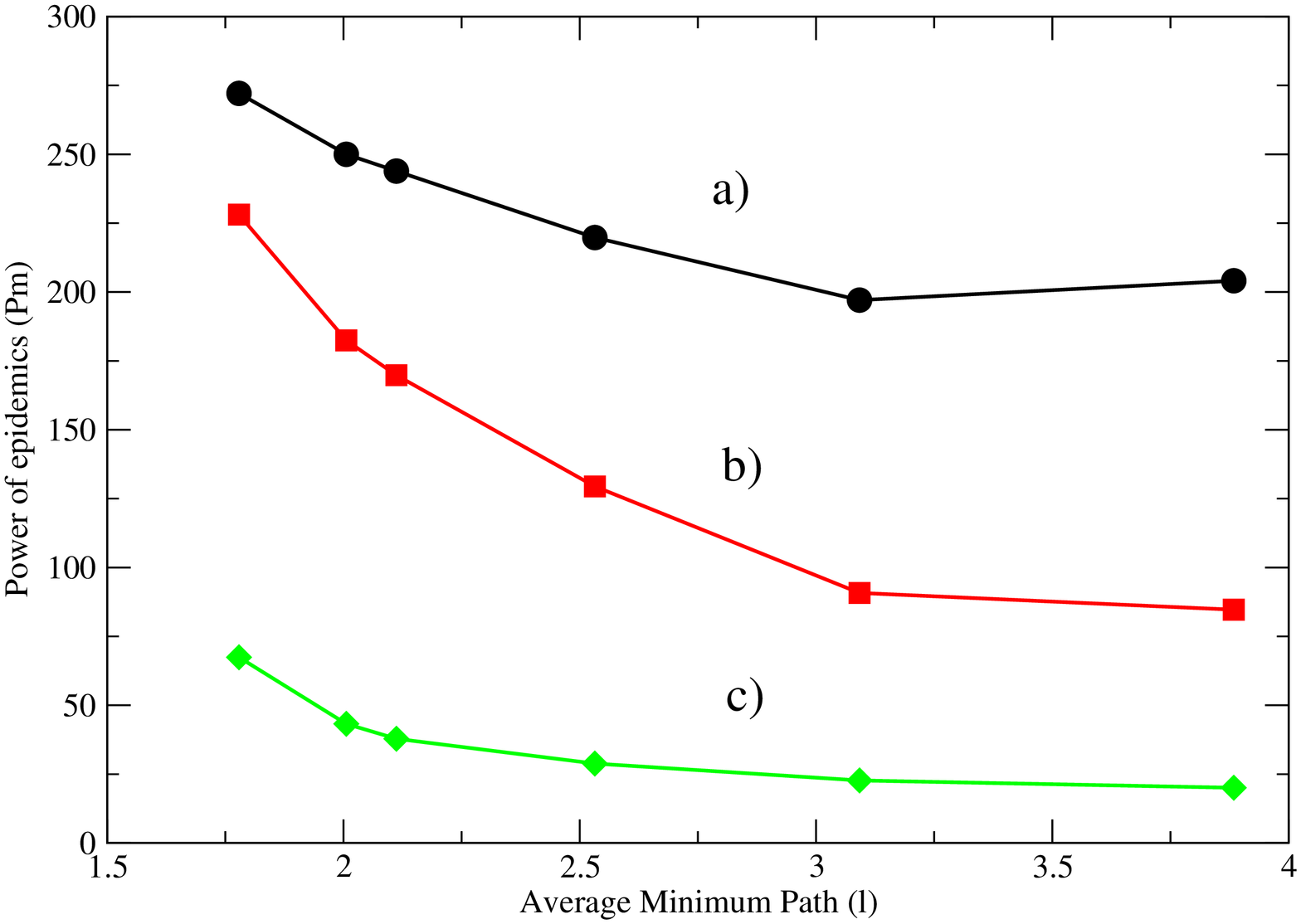}
\caption{Comparison of the mean power of the epidemics for the different
patterns of human mobility (from left to right: Random, Levy1.65, Levy2, Levy3, Levy4, Move400), characterized by the average minimum
path for three conditions: $(a)$ 400 BS and constant
temperature of 23 degrees and $(b)$ 400 BS and $(c)$ 200 BS for seasonal
variation of temperature.}
\label{marce5}
\end{figure}

Figure \ref{marce5} shows the mean power for three conditions: $(a)$ 400 BS
and constant temperature of 23 degrees, $(b)$ 400 BS and seasonal variation
of temperature and $(c)$ 200 BS and seasonal variation of temperature. 
The $P_{m}$ grows with broader jump length distributions and with higher BS densities.
For the case of 400 BS is higher for constant temperature
than for seasonal variation of temperature. If we compare 400 BS and 200 BS
(for seasonal variation of temperature) we see the same increasing tendency
of $P_{m}$ with human mobility but this value is higher for 400
BS than for 200 BS. (For the case of 400 BS and constant T the increase of the
power of the epidemic is a consequence of the reduction of its duration as the
pattern of human mobility approaches the fully random case. At constant T all of the
population gets infected. On the other hand, when the Temperature is not fixed it severely
constrains the mosquitoes population and then the increase in $P_{m}$ is mainly due to the
increase of the size of the infected population as the time span of the epidemics is only mildly 
dependent on the driving force of the dispersal. 

\bigskip

\section{Conclusions}
In this work we have explored the effect of human mobility on the dynamics of a vector borne infection.
We have added this characteristic of human behavior on an already tested model of dengue dispersal when the dynamics is driven by mosquitoes alone.
We have analyzed the case of a schematic city of 20x20 blocks with 100 individuals per block.

We have considered two temperature profiles, on the one hand a simple constant temperature one and a 
realistic time distribution corresponding to the city of Buenos Aires, Argentina.

Another variable in our analysis has been the number of breeding cites in the city, we have considered 
50,100, 200, 300 and 400 breeding cites per block.  

Human mobility has been described by superimposing diverse kinds of networks in which links represent
the daily movement of humans. The distribution of lengths of this links are derived from recent 
studies on human motion and in particular, taking into account the finding that human behavior is 
highly predictable. We have also considered reference patterns i.e purely random motion and random 
motion of a single human per block.

We have explored different observables like size and duration of the outbreaks, and complementary the 
morphological characteristics of the pattern of recovered individuals.

We have found that human mobility strongly enhaces the infection dispersal. Even for the case in 
which just one individual per block can perform a long jump. This effect can be traced to the fact that 
when the disease dispersal is driven by mosquitoes alone we have a single focus that expands due to 
diffusive kind of diynamics. When human mobility is taken into account, multiple loci appear as the 
time evolution is followed 

Human mobility increases the size and the speed of propagation of the outbreaks. This feature 
can be captured by the magnitude "Power of the epidemics" defined as the quotient of the size of the epidemics 
divided by the its time span. This magnitude displays a monotonous increase as the Mean Length Path of the 
network describing the daily human mobility pattern decreases.

This findings indicate that human mobility might turn out to be the main driving force in the epidemics 
dynamics.

Both in the case of fixed temperature and seasonal variational one, human motion gives rise to faster 
and more widespread epidemics.

Finally this findings indicate that, when considering meassures to fight epidemics dispersal human motion should be one of the top 
concerns. We are presently exploring this issue.

Acknowledgments

C.O.D, M.O and H.G.S are members of the Carrera del Investigador CONICET. D.H.B is a fellow of the 
CONICET.
We thank the support by the University of Buenos Aires (grant X210).

\bibliography{referencias}

\begin{thebibliography}{43}
\expandafter\ifx\csname natexlab\endcsname\relax\def\natexlab#1{#1}\fi
\expandafter\ifx\csname bibnamefont\endcsname\relax
  \def\bibnamefont#1{#1}\fi
\expandafter\ifx\csname bibfnamefont\endcsname\relax
  \def\bibfnamefont#1{#1}\fi
\expandafter\ifx\csname citenamefont\endcsname\relax
  \def\citenamefont#1{#1}\fi
\expandafter\ifx\csname url\endcsname\relax
  \def\url#1{\texttt{#1}}\fi
\expandafter\ifx\csname urlprefix\endcsname\relax\def\urlprefix{URL }\fi
\providecommand{\bibinfo}[2]{#2}
\providecommand{\eprint}[2][]{\url{#2}}

\bibitem[{\citenamefont{Gubler}(1998)}]{gubl98}
\bibinfo{author}{\bibfnamefont{D.~J.} \bibnamefont{Gubler}},
  \bibinfo{journal}{Clinical Microbiology Review}
  \textbf{\bibinfo{volume}{11}}, \bibinfo{pages}{480} (\bibinfo{year}{1998}).

\bibitem[{\citenamefont{Newton and Reiter}(1992)}]{newt92}
\bibinfo{author}{\bibfnamefont{E.~A.~C.} \bibnamefont{Newton}}
  \bibnamefont{and} \bibinfo{author}{\bibfnamefont{P.}~\bibnamefont{Reiter}},
  \bibinfo{journal}{Am. J. Trop. Med. Hyg.} \textbf{\bibinfo{volume}{47}},
  \bibinfo{pages}{709} (\bibinfo{year}{1992}).

\bibitem[{\citenamefont{Focks et~al.}(1993{\natexlab{a}})\citenamefont{Focks,
  Haile, Daniels, and Moun}}]{fock93a}
\bibinfo{author}{\bibfnamefont{D.~A.} \bibnamefont{Focks}},
  \bibinfo{author}{\bibfnamefont{D.~C.} \bibnamefont{Haile}},
  \bibinfo{author}{\bibfnamefont{E.}~\bibnamefont{Daniels}}, \bibnamefont{and}
  \bibinfo{author}{\bibfnamefont{G.~A.} \bibnamefont{Moun}},
  \bibinfo{journal}{Journal of Medical Entomology}
  \textbf{\bibinfo{volume}{30}}, \bibinfo{pages}{1003}
  (\bibinfo{year}{1993}{\natexlab{a}}).

\bibitem[{\citenamefont{Focks et~al.}(1993{\natexlab{b}})\citenamefont{Focks,
  Haile, Daniels, and Mount}}]{fock93b}
\bibinfo{author}{\bibfnamefont{D.~A.} \bibnamefont{Focks}},
  \bibinfo{author}{\bibfnamefont{D.~C.} \bibnamefont{Haile}},
  \bibinfo{author}{\bibfnamefont{E.}~\bibnamefont{Daniels}}, \bibnamefont{and}
  \bibinfo{author}{\bibfnamefont{G.~A.} \bibnamefont{Mount}},
  \bibinfo{journal}{Journal of Medical Entomology}
  \textbf{\bibinfo{volume}{30}}, \bibinfo{pages}{1019}
  (\bibinfo{year}{1993}{\natexlab{b}}).

\bibitem[{\citenamefont{Focks et~al.}(1995)\citenamefont{Focks, Haile, Daniels,
  and Keesling}}]{fock95}
\bibinfo{author}{\bibfnamefont{D.~A.} \bibnamefont{Focks}},
  \bibinfo{author}{\bibfnamefont{D.~C.} \bibnamefont{Haile}},
  \bibinfo{author}{\bibfnamefont{E.}~\bibnamefont{Daniels}}, \bibnamefont{and}
  \bibinfo{author}{\bibfnamefont{D.}~\bibnamefont{Keesling}},
  \bibinfo{journal}{Am. J. Trop. Med. Hyg.} \textbf{\bibinfo{volume}{53}},
  \bibinfo{pages}{489} (\bibinfo{year}{1995}).

\bibitem[{\citenamefont{Otero and Solari}(2010)}]{oter10}
\bibinfo{author}{\bibfnamefont{M.}~\bibnamefont{Otero}} \bibnamefont{and}
  \bibinfo{author}{\bibfnamefont{H.~G.} \bibnamefont{Solari}},
  \bibinfo{journal}{Mathematical Biosciences} \textbf{\bibinfo{volume}{223}},
  \bibinfo{pages}{32} (\bibinfo{year}{2010}).

\bibitem[{\citenamefont{Otero et~al.}(2006)\citenamefont{Otero, Solari, and
  Schweigmann}}]{oter06}
\bibinfo{author}{\bibfnamefont{M.}~\bibnamefont{Otero}},
  \bibinfo{author}{\bibfnamefont{H.~G.} \bibnamefont{Solari}},
  \bibnamefont{and}
  \bibinfo{author}{\bibfnamefont{N.}~\bibnamefont{Schweigmann}},
  \bibinfo{journal}{Bull. Math. Biol.} \textbf{\bibinfo{volume}{68}},
  \bibinfo{pages}{1945} (\bibinfo{year}{2006}).

\bibitem[{\citenamefont{Otero et~al.}(2008)\citenamefont{Otero, Schweigmann,
  and Solari}}]{oter08}
\bibinfo{author}{\bibfnamefont{M.}~\bibnamefont{Otero}},
  \bibinfo{author}{\bibfnamefont{N.}~\bibnamefont{Schweigmann}},
  \bibnamefont{and} \bibinfo{author}{\bibfnamefont{H.~G.}
  \bibnamefont{Solari}}, \bibinfo{journal}{Bulletin of Mathematical Biology}
  \textbf{\bibinfo{volume}{70}}, \bibinfo{pages}{1297} (\bibinfo{year}{2008}).

\bibitem[{\citenamefont{Esteva and Vargas}(1998)}]{este98}
\bibinfo{author}{\bibfnamefont{L.}~\bibnamefont{Esteva}} \bibnamefont{and}
  \bibinfo{author}{\bibfnamefont{C.}~\bibnamefont{Vargas}},
  \bibinfo{journal}{Mathematical Biosciences} \textbf{\bibinfo{volume}{150}},
  \bibinfo{pages}{131} (\bibinfo{year}{1998}).

\bibitem[{\citenamefont{Esteva and Vargas}(1999)}]{este99}
\bibinfo{author}{\bibfnamefont{L.}~\bibnamefont{Esteva}} \bibnamefont{and}
  \bibinfo{author}{\bibfnamefont{C.}~\bibnamefont{Vargas}},
  \bibinfo{journal}{Journal of Mathematical Biology}
  \textbf{\bibinfo{volume}{38}}, \bibinfo{pages}{220} (\bibinfo{year}{1999}).

\bibitem[{\citenamefont{Esteva and Vargas}(2000)}]{este00}
\bibinfo{author}{\bibfnamefont{L.}~\bibnamefont{Esteva}} \bibnamefont{and}
  \bibinfo{author}{\bibfnamefont{C.}~\bibnamefont{Vargas}},
  \bibinfo{journal}{Mathematical Biosciences} \textbf{\bibinfo{volume}{167}},
  \bibinfo{pages}{51} (\bibinfo{year}{2000}).

\bibitem[{\citenamefont{Bartley et~al.}(2002)\citenamefont{Bartley, Donnelly,
  and Garnett}}]{bart02}
\bibinfo{author}{\bibfnamefont{L.~M.} \bibnamefont{Bartley}},
  \bibinfo{author}{\bibfnamefont{C.~A.} \bibnamefont{Donnelly}},
  \bibnamefont{and} \bibinfo{author}{\bibfnamefont{G.~P.}
  \bibnamefont{Garnett}}, \bibinfo{journal}{Transactions of the royal society
  of tropical medicine and hygiene} \textbf{\bibinfo{volume}{96}},
  \bibinfo{pages}{387} (\bibinfo{year}{2002}).

\bibitem[{\citenamefont{Pongsumpun and Tang}(2003)}]{pong03}
\bibinfo{author}{\bibfnamefont{P.}~\bibnamefont{Pongsumpun}} \bibnamefont{and}
  \bibinfo{author}{\bibfnamefont{I.~M.} \bibnamefont{Tang}},
  \bibinfo{journal}{Mathematical and Computer Modelling}
  \textbf{\bibinfo{volume}{37}}, \bibinfo{pages}{949} (\bibinfo{year}{2003}).

\bibitem[{\citenamefont{Chowella et~al.}(2007)\citenamefont{Chowella,
  Diaz-Dueñas, Miller, Alcazar-Velazco, Hyman, Fenimore, and
  Castillo-Chavez}}]{chow07}
\bibinfo{author}{\bibfnamefont{G.}~\bibnamefont{Chowella}},
  \bibinfo{author}{\bibfnamefont{P.}~\bibnamefont{Diaz-Dueñas}},
  \bibinfo{author}{\bibfnamefont{J.}~\bibnamefont{Miller}},
  \bibinfo{author}{\bibfnamefont{A.}~\bibnamefont{Alcazar-Velazco}},
  \bibinfo{author}{\bibfnamefont{J.}~\bibnamefont{Hyman}},
  \bibinfo{author}{\bibfnamefont{P.}~\bibnamefont{Fenimore}}, \bibnamefont{and}
  \bibinfo{author}{\bibfnamefont{C.}~\bibnamefont{Castillo-Chavez}},
  \bibinfo{journal}{Mathematical Biosciences} \textbf{\bibinfo{volume}{208}},
  \bibinfo{pages}{571} (\bibinfo{year}{2007}).

\bibitem[{\citenamefont{Favier et~al.}(2005)\citenamefont{Favier, Schmit,
  Müller-Graf, Cazelles, Degallier, Mondet, and Dubois}}]{favi05}
\bibinfo{author}{\bibfnamefont{C.}~\bibnamefont{Favier}},
  \bibinfo{author}{\bibfnamefont{D.}~\bibnamefont{Schmit}},
  \bibinfo{author}{\bibfnamefont{C.~D.~M.} \bibnamefont{Müller-Graf}},
  \bibinfo{author}{\bibfnamefont{B.}~\bibnamefont{Cazelles}},
  \bibinfo{author}{\bibfnamefont{N.}~\bibnamefont{Degallier}},
  \bibinfo{author}{\bibfnamefont{B.}~\bibnamefont{Mondet}}, \bibnamefont{and}
  \bibinfo{author}{\bibfnamefont{M.~A.} \bibnamefont{Dubois}},
  \bibinfo{journal}{Proceedings of the Royal Society (London): Biological
  Sciences} \textbf{\bibinfo{volume}{272}}, \bibinfo{pages}{1171}
  (\bibinfo{year}{2005}).

\bibitem[{\citenamefont{Otero et~al.}(2010)\citenamefont{Otero, Barmak, Dorso,
  Solari, and Natiello}}]{oter11}
\bibinfo{author}{\bibfnamefont{M.}~\bibnamefont{Otero}},
  \bibinfo{author}{\bibfnamefont{D.}~\bibnamefont{Barmak}},
  \bibinfo{author}{\bibfnamefont{C.}~\bibnamefont{Dorso}},
  \bibinfo{author}{\bibfnamefont{H.}~\bibnamefont{Solari}}, \bibnamefont{and}
  \bibinfo{author}{\bibfnamefont{M.}~\bibnamefont{Natiello}}
  (\bibinfo{year}{2010}), \bibinfo{note}{arXiv:1012.1281v1 [q-bio.PE]}.

\bibitem[{\citenamefont{Epstein et~al.}(2008)\citenamefont{Epstein, Parker,
  Cummings, and Hammond}}]{Epst08}
\bibinfo{author}{\bibfnamefont{J.~M.} \bibnamefont{Epstein}},
  \bibinfo{author}{\bibfnamefont{J.}~\bibnamefont{Parker}},
  \bibinfo{author}{\bibfnamefont{D.}~\bibnamefont{Cummings}}, \bibnamefont{and}
  \bibinfo{author}{\bibfnamefont{R.~A.} \bibnamefont{Hammond}},
  \bibinfo{journal}{PLoS ONE} \textbf{\bibinfo{volume}{3}},
  \bibinfo{pages}{e3955} (\bibinfo{year}{2008}),
  \urlprefix\url{http://dx.doi.org/10.1371%2Fjournal.pone.0003955}.

\bibitem[{\citenamefont{Gross and Blasius}(2008)}]{Gros08}
\bibinfo{author}{\bibfnamefont{T.}~\bibnamefont{Gross}} \bibnamefont{and}
  \bibinfo{author}{\bibfnamefont{B.}~\bibnamefont{Blasius}},
  \bibinfo{journal}{J. R. Soc. Interface} \textbf{\bibinfo{volume}{5}},
  \bibinfo{pages}{259} (\bibinfo{year}{2008}),
  \urlprefix\url{http://dx.doi.org/10.1098/rsif.2007.1229}.

\bibitem[{\citenamefont{Gross et~al.}(2006)\citenamefont{Gross, D'Lima, and
  Blasius}}]{Gros06}
\bibinfo{author}{\bibfnamefont{T.}~\bibnamefont{Gross}},
  \bibinfo{author}{\bibfnamefont{C.~J.} \bibnamefont{D'Lima}},
  \bibnamefont{and} \bibinfo{author}{\bibfnamefont{B.}~\bibnamefont{Blasius}},
  \bibinfo{journal}{Physical Review Letters} \textbf{\bibinfo{volume}{96}},
  \bibinfo{pages}{208701+} (\bibinfo{year}{2006}),
  \urlprefix\url{http://dx.doi.org/10.1103/PhysRevLett.96.208701}.

\bibitem[{\citenamefont{Risau-Gusmans et~al.}(2009)\citenamefont{Risau-Gusmans,
  , and Zanette}}]{Risa09}
\bibinfo{author}{\bibfnamefont{S.}~\bibnamefont{Risau-Gusmans}}, ,
  \bibnamefont{and} \bibinfo{author}{\bibfnamefont{D.~H.}
  \bibnamefont{Zanette}}, \bibinfo{journal}{Journal of Theoretical Biology}
  \textbf{\bibinfo{volume}{257}}, \bibinfo{pages}{52} (\bibinfo{year}{2009}),
  ISSN \bibinfo{issn}{00225193},
  \urlprefix\url{http://dx.doi.org/10.1016/j.jtbi.2008.10.027}.

\bibitem[{\citenamefont{Zanette}(2007)}]{Zane07}
\bibinfo{author}{\bibfnamefont{D.~H.} \bibnamefont{Zanette}}
  (\bibinfo{year}{2007}), \bibinfo{note}{arXiv e-prints}, \eprint{0707.1249},
  \urlprefix\url{http://arxiv.org/abs/0707.1249}.

\bibitem[{\citenamefont{Zanette and Gusman}(2007)}]{Zane07-2}
\bibinfo{author}{\bibfnamefont{D.~H.} \bibnamefont{Zanette}} \bibnamefont{and}
  \bibinfo{author}{\bibfnamefont{S.~R.} \bibnamefont{Gusman}},
  \bibinfo{journal}{Journal of Biological Physics}
  \textbf{\bibinfo{volume}{34}}, \bibinfo{pages}{135} (\bibinfo{year}{2007}),
  \eprint{0711.0874}, \urlprefix\url{http://arxiv.org/abs/0711.0874}.

\bibitem[{\citenamefont{Lig et~al.}(2007)\citenamefont{Lig, Jia-Ren, Jian-Guo,
  and Zi-Ran}}]{Wang07}
\bibinfo{author}{\bibfnamefont{W.}~\bibnamefont{Lig}},
  \bibinfo{author}{\bibfnamefont{Y.}~\bibnamefont{Jia-Ren}},
  \bibinfo{author}{\bibfnamefont{Z.}~\bibnamefont{Jian-Guo}}, \bibnamefont{and}
  \bibinfo{author}{\bibfnamefont{L.}~\bibnamefont{Zi-Ran}},
  \bibinfo{journal}{Chinese Physics} \textbf{\bibinfo{volume}{16}},
  \bibinfo{pages}{2498+} (\bibinfo{year}{2007}), ISSN
  \bibinfo{issn}{1009-1963},
  \urlprefix\url{http://dx.doi.org/10.1088/1009-1963/16/9/002}.

\bibitem[{\citenamefont{Fefferman and Ng}(2007)}]{Feff07}
\bibinfo{author}{\bibfnamefont{N.~H.} \bibnamefont{Fefferman}}
  \bibnamefont{and} \bibinfo{author}{\bibfnamefont{K.~L.} \bibnamefont{Ng}},
  \bibinfo{journal}{Physical Review E} \textbf{\bibinfo{volume}{76}},
  \bibinfo{pages}{031919+} (\bibinfo{year}{2007}),
  \urlprefix\url{http://dx.doi.org/10.1103/PhysRevE.76.031919}.

\bibitem[{\citenamefont{Funk et~al.}(2010)\citenamefont{Funk, Salath\'{e}, and
  Jansen}}]{Funk10}
\bibinfo{author}{\bibfnamefont{S.}~\bibnamefont{Funk}},
  \bibinfo{author}{\bibfnamefont{M.}~\bibnamefont{Salath\'{e}}},
  \bibnamefont{and} \bibinfo{author}{\bibfnamefont{V.~A.~A.}
  \bibnamefont{Jansen}}, \bibinfo{journal}{Journal of The Royal Society
  Interface}  (\bibinfo{year}{2010}),
  \urlprefix\url{http://dx.doi.org/10.1098/rsif.2010.0142}.

\bibitem[{\citenamefont{Zhao et~al.}(2010)\citenamefont{Zhao, Calder\'{o}n, Xu,
  Zao, Fenn, Sornette, Crane, Hui, and Johnson}}]{Zhao10}
\bibinfo{author}{\bibfnamefont{Z.}~\bibnamefont{Zhao}},
  \bibinfo{author}{\bibfnamefont{J.~P.} \bibnamefont{Calder\'{o}n}},
  \bibinfo{author}{\bibfnamefont{C.}~\bibnamefont{Xu}},
  \bibinfo{author}{\bibfnamefont{G.}~\bibnamefont{Zao}},
  \bibinfo{author}{\bibfnamefont{D.}~\bibnamefont{Fenn}},
  \bibinfo{author}{\bibfnamefont{D.}~\bibnamefont{Sornette}},
  \bibinfo{author}{\bibfnamefont{R.}~\bibnamefont{Crane}},
  \bibinfo{author}{\bibfnamefont{P.~M.} \bibnamefont{Hui}}, \bibnamefont{and}
  \bibinfo{author}{\bibfnamefont{N.~F.} \bibnamefont{Johnson}},
  \bibinfo{journal}{Physical Review E} \textbf{\bibinfo{volume}{81}},
  \bibinfo{pages}{056107+} (\bibinfo{year}{2010}),
  \urlprefix\url{http://dx.doi.org/10.1103/PhysRevE.81.056107}.

\bibitem[{\citenamefont{Sattenspiel}(2009)}]{Lisa09}
\bibinfo{author}{\bibfnamefont{L.}~\bibnamefont{Sattenspiel}},
  \emph{\bibinfo{title}{The Geographic Spread of Infectious Diseases:Models and
  Applications}} (\bibinfo{publisher}{Princeton University},
  \bibinfo{year}{2009}).

\bibitem[{\citenamefont{Rhee et~al.}(2007)\citenamefont{Rhee, Shin, Hong, Lee,
  and Chong}}]{injo07}
\bibinfo{author}{\bibfnamefont{I.}~\bibnamefont{Rhee}},
  \bibinfo{author}{\bibfnamefont{M.}~\bibnamefont{Shin}},
  \bibinfo{author}{\bibfnamefont{S.}~\bibnamefont{Hong}},
  \bibinfo{author}{\bibfnamefont{K.}~\bibnamefont{Lee}}, \bibnamefont{and}
  \bibinfo{author}{\bibfnamefont{S.}~\bibnamefont{Chong}},
  \bibinfo{journal}{Technical Report, Computer Science Department, North
  Carolina State University}  (\bibinfo{year}{2007}).

\bibitem[{\citenamefont{Gonzalez et~al.}(2008)\citenamefont{Gonzalez, Hidalgo,
  and Barabasi}}]{bara08}
\bibinfo{author}{\bibfnamefont{M.~C.} \bibnamefont{Gonzalez}},
  \bibinfo{author}{\bibfnamefont{C.~A.} \bibnamefont{Hidalgo}},
  \bibnamefont{and} \bibinfo{author}{\bibfnamefont{A.-L.}
  \bibnamefont{Barabasi}}, \bibinfo{journal}{Nature}
  \textbf{\bibinfo{volume}{453}}, \bibinfo{pages}{779} (\bibinfo{year}{2008}).

\bibitem[{\citenamefont{Brockmann et~al.}(2006)\citenamefont{Brockmann,
  Hufnagel, and Geisel}}]{Broc06}
\bibinfo{author}{\bibfnamefont{D.}~\bibnamefont{Brockmann}},
  \bibinfo{author}{\bibfnamefont{L.}~\bibnamefont{Hufnagel}}, \bibnamefont{and}
  \bibinfo{author}{\bibfnamefont{T.}~\bibnamefont{Geisel}},
  \bibinfo{journal}{Nature} \textbf{\bibinfo{volume}{439}}
  (\bibinfo{year}{2006}).

\bibitem[{\citenamefont{D.~Brockmann}(2007)}]{Broc07}
\bibinfo{author}{\bibfnamefont{L.~H.} \bibnamefont{D.~Brockmann}},
  \emph{\bibinfo{title}{The scaling law of human travel - A message from
  George}} (\bibinfo{publisher}{World Scientific}, \bibinfo{year}{2007}).

\bibitem[{\citenamefont{Chowell et~al.}(2003)\citenamefont{Chowell, Hyman,
  Eubank, and Castillo-Chavez}}]{Chow03}
\bibinfo{author}{\bibfnamefont{G.}~\bibnamefont{Chowell}},
  \bibinfo{author}{\bibfnamefont{J.~M.} \bibnamefont{Hyman}},
  \bibinfo{author}{\bibfnamefont{S.}~\bibnamefont{Eubank}}, \bibnamefont{and}
  \bibinfo{author}{\bibfnamefont{C.}~\bibnamefont{Castillo-Chavez}},
  \bibinfo{journal}{Physical Review E} \textbf{\bibinfo{volume}{68}},
  \bibinfo{pages}{661021} (\bibinfo{year}{2003}).

\bibitem[{\citenamefont{Cattuto et~al.}(2010)\citenamefont{Cattuto, Van~den
  Broeck, Barrat, Colizza, Pinton, and Vespignani}}]{Catt10}
\bibinfo{author}{\bibfnamefont{C.}~\bibnamefont{Cattuto}},
  \bibinfo{author}{\bibfnamefont{W.}~\bibnamefont{Van~den Broeck}},
  \bibinfo{author}{\bibfnamefont{A.}~\bibnamefont{Barrat}},
  \bibinfo{author}{\bibfnamefont{V.}~\bibnamefont{Colizza}},
  \bibinfo{author}{\bibfnamefont{J.-F.} \bibnamefont{Pinton}},
  \bibnamefont{and}
  \bibinfo{author}{\bibfnamefont{A.}~\bibnamefont{Vespignani}},
  \bibinfo{journal}{PLoS ONE} \textbf{\bibinfo{volume}{5}},
  \bibinfo{pages}{e11596} (\bibinfo{year}{2010}),
  \urlprefix\url{http://dx.doi.org/10.1371%2Fjournal.pone.0011596}.

\bibitem[{\citenamefont{Candia et~al.}(2007)\citenamefont{Candia, Gonzalez,
  Wang, and Schoenharl}}]{Cand07}
\bibinfo{author}{\bibfnamefont{J.}~\bibnamefont{Candia}},
  \bibinfo{author}{\bibfnamefont{M.~C.} \bibnamefont{Gonzalez}},
  \bibinfo{author}{\bibfnamefont{P.}~\bibnamefont{Wang}}, \bibnamefont{and}
  \bibinfo{author}{\bibfnamefont{T.}~\bibnamefont{Schoenharl}}
  (\bibinfo{year}{2007}), \bibinfo{note}{detection of anomalous local cell
  pattern}, \urlprefix\url{http://arxiv.org/pdf/0710.2939}.

\bibitem[{\citenamefont{Wang and Gonzalez}(2009)}]{Gonz09}
\bibinfo{author}{\bibfnamefont{P.}~\bibnamefont{Wang}} \bibnamefont{and}
  \bibinfo{author}{\bibfnamefont{M.~C.} \bibnamefont{Gonzalez}},
  \bibinfo{journal}{Royal Society of London Philosophical Transactions Series
  A} \textbf{\bibinfo{volume}{367}}, \bibinfo{pages}{3321}
  (\bibinfo{year}{2009}).

\bibitem[{\citenamefont{Buscarino et~al.}(2008)\citenamefont{Buscarino,
  Fortuna, Frasca, and Latora}}]{Lato08}
\bibinfo{author}{\bibfnamefont{A.}~\bibnamefont{Buscarino}},
  \bibinfo{author}{\bibfnamefont{L.}~\bibnamefont{Fortuna}},
  \bibinfo{author}{\bibfnamefont{M.}~\bibnamefont{Frasca}}, \bibnamefont{and}
  \bibinfo{author}{\bibfnamefont{V.}~\bibnamefont{Latora}},
  \bibinfo{journal}{European Physics Letters} \textbf{\bibinfo{volume}{82}},
  \bibinfo{pages}{38002} (\bibinfo{year}{2008}).

\bibitem[{\citenamefont{Li et~al.}(2010)\citenamefont{Li, Cao, and
  Cao}}]{Li110}
\bibinfo{author}{\bibfnamefont{X.}~\bibnamefont{Li}},
  \bibinfo{author}{\bibfnamefont{L.}~\bibnamefont{Cao}}, \bibnamefont{and}
  \bibinfo{author}{\bibfnamefont{G.~F.} \bibnamefont{Cao}},
  \bibinfo{journal}{The European Physical Journal B - Condensed Matter and
  Complex Systems}  (\bibinfo{year}{2010}), ISSN \bibinfo{issn}{1434-6028},
  \urlprefix\url{http://dx.doi.org/10.1140/epjb/e2010-00090-9}.

\bibitem[{\citenamefont{Keeling and Eames}(2005)}]{Keel07}
\bibinfo{author}{\bibfnamefont{M.}~\bibnamefont{Keeling}} \bibnamefont{and}
  \bibinfo{author}{\bibfnamefont{K.~T.} \bibnamefont{Eames}},
  \bibinfo{journal}{Journal of the Royal Society Interface 22}
  \textbf{\bibinfo{volume}{2}}, \bibinfo{pages}{295} (\bibinfo{year}{2005}),
  ISSN \bibinfo{issn}{1742-5689},
  \urlprefix\url{http://dx.doi.org/10.1098/rsif.2005.0051}.

\bibitem[{\citenamefont{Pongsumpun et~al.}(2008)\citenamefont{Pongsumpun,
  Lopez, Favier, Torres, Llosa, and Dubois}}]{Pong08}
\bibinfo{author}{\bibfnamefont{P.}~\bibnamefont{Pongsumpun}},
  \bibinfo{author}{\bibfnamefont{D.~G.} \bibnamefont{Lopez}},
  \bibinfo{author}{\bibfnamefont{C.}~\bibnamefont{Favier}},
  \bibinfo{author}{\bibfnamefont{L.}~\bibnamefont{Torres}},
  \bibinfo{author}{\bibfnamefont{J.}~\bibnamefont{Llosa}}, \bibnamefont{and}
  \bibinfo{author}{\bibfnamefont{M.}~\bibnamefont{Dubois}},
  \bibinfo{journal}{Tropical Medicine and International Health}
  \textbf{\bibinfo{volume}{13}}, \bibinfo{pages}{1180} (\bibinfo{year}{2008}).

\bibitem[{\citenamefont{Stoddard et~al.}(2009)\citenamefont{Stoddard, Morrison,
  Vazquez-Prokopec, Soldan, Kochel, Kitron, Elder, and Scott}}]{stod09}
\bibinfo{author}{\bibfnamefont{S.~T.} \bibnamefont{Stoddard}},
  \bibinfo{author}{\bibfnamefont{A.~C.} \bibnamefont{Morrison}},
  \bibinfo{author}{\bibfnamefont{G.~M.} \bibnamefont{Vazquez-Prokopec}},
  \bibinfo{author}{\bibfnamefont{V.~P.} \bibnamefont{Soldan}},
  \bibinfo{author}{\bibfnamefont{T.~J.} \bibnamefont{Kochel}},
  \bibinfo{author}{\bibfnamefont{U.}~\bibnamefont{Kitron}},
  \bibinfo{author}{\bibfnamefont{J.~P.} \bibnamefont{Elder}}, \bibnamefont{and}
  \bibinfo{author}{\bibfnamefont{T.~W.} \bibnamefont{Scott}},
  \bibinfo{journal}{PLoS Negl Trop Dis} \textbf{\bibinfo{volume}{3}},
  \bibinfo{pages}{e481} (\bibinfo{year}{2009}),
  \urlprefix\url{http://dx.doi.org/10.1371/journal.pntd.0000481}.

\bibitem[{\citenamefont{Seijo et~al.}(2009)\citenamefont{Seijo, Romer,
  Espinosa, Monroig, Giamperetti, Ameri, and Antonelli}}]{seij09}
\bibinfo{author}{\bibfnamefont{A.}~\bibnamefont{Seijo}},
  \bibinfo{author}{\bibfnamefont{Y.}~\bibnamefont{Romer}},
  \bibinfo{author}{\bibfnamefont{M.}~\bibnamefont{Espinosa}},
  \bibinfo{author}{\bibfnamefont{J.}~\bibnamefont{Monroig}},
  \bibinfo{author}{\bibfnamefont{S.}~\bibnamefont{Giamperetti}},
  \bibinfo{author}{\bibfnamefont{D.}~\bibnamefont{Ameri}}, \bibnamefont{and}
  \bibinfo{author}{\bibfnamefont{L.}~\bibnamefont{Antonelli}},
  \bibinfo{journal}{Medicina} \textbf{\bibinfo{volume}{69}},
  \bibinfo{pages}{593} (\bibinfo{year}{2009}), \bibinfo{note}{iSSN 0025-7680}.

\bibitem[{\citenamefont{Nishiura and Halstead}(2007)}]{nish07}
\bibinfo{author}{\bibfnamefont{H.}~\bibnamefont{Nishiura}} \bibnamefont{and}
  \bibinfo{author}{\bibfnamefont{S.~B.} \bibnamefont{Halstead}},
  \bibinfo{journal}{Journal of Infectious Diseases}
  \textbf{\bibinfo{volume}{195}}, \bibinfo{pages}{1007} (\bibinfo{year}{2007}).

\bibitem[{\citenamefont{Song et~al.}(2010)\citenamefont{Song, Qu, Blumm, and
  Barabási}}]{Chao10}
\bibinfo{author}{\bibfnamefont{C.}~\bibnamefont{Song}},
  \bibinfo{author}{\bibfnamefont{Z.}~\bibnamefont{Qu}},
  \bibinfo{author}{\bibfnamefont{N.}~\bibnamefont{Blumm}}, \bibnamefont{and}
  \bibinfo{author}{\bibfnamefont{A.-L.} \bibnamefont{Barabási}},
  \bibinfo{journal}{Science} \textbf{\bibinfo{volume}{327}},
  \bibinfo{pages}{1018} (\bibinfo{year}{2010}),
  \eprint{http://www.sciencemag.org/cgi/reprint/327/5968/1018.pdf},
  \urlprefix\url{http://www.barabasilab.com/pubs/CCNR-ALB_Publications/201002-%
19_Science-Predictability/201002-19_Science-Predictability.pdf}.

\end{thebibliography}


\end{document}